\newif\ifShowKeys
\ShowKeystrue
\documentclass[11pt]{article}
\pdfoutput=1
\usepackage[usenames,dvipsnames]{xcolor}
\usepackage[setpagesize=false,pagebackref=false, linktocpage, bookmarksopen=true, colorlinks=true, linkcolor=Maroon,citecolor=Maroon,urlcolor=Maroon]{hyperref}
 \usepackage[parsep]{collref}

\usepackage{tocloft}

\usepackage{amsmath, amssymb,amsthm}
\usepackage{amsthm}
\numberwithin{equation}{section}

\usepackage{tabto}
\usepackage{bm,environ,mathrsfs,array,arydshln}
\usepackage{booktabs,float,slashed,hyperref}
\usepackage[mathcal]{euscript}
\usepackage{tensor} 						% Ratcliffe package to write tensors
\usepackage{mathabx}
\usepackage{simpler-wick}
\usepackage[vcentermath]{youngtab}

\usepackage{aurical}
\usepackage[T1]{fontenc}
\usepackage[nodayofweek]{date time}
\usepackage{graphicx,epsfig,epic}

\usepackage{framed}						% for shaded equations \begin{shaded}...\end{shaded}
\definecolor{shadecolor}{rgb}{0.9996078, 0.984314, 0.960784}

\usepackage{comment}

\allowdisplaybreaks

\definecolor{myred}{RGB}{233, 33, 45}

\newcommand{\bs}{\begin{shaded}}
\newcommand{\es}{\end{shaded}\noindent}
\def\ba#1\ea{\begin{align}#1\end{align}}		% very clever way to bypass the known problem...
\newcommand{\be}{\begin{equation}}
\newcommand{\ee}{\end{equation}}
\newcommand{\mc}{\mathcal }

\newcommand{\la}{\label}
\newcommand{\eps}{\varepsilon}

\newcommand{\lp}{\notag \\ & }

\newcommand{\wt}{\widetilde}

\newcommand{\cf}{\textit{cf.} }

\newcommand{\N}{\mathcal N}

\makeatletter
\DeclareFontFamily{OMX}{MnSymbolE}{}
\DeclareSymbolFont{MnLargeSymbols}{OMX}{MnSymbolE}{m}{n}
\SetSymbolFont{MnLargeSymbols}{bold}{OMX}{MnSymbolE}{b}{n}
\DeclareFontShape{OMX}{MnSymbolE}{m}{n}{
<-6>  MnSymbolE5
   <6-7>  MnSymbolE6
   <7-8>  MnSymbolE7
   <8-9>  MnSymbolE8
   <9-10> MnSymbolE9
  <10-12> MnSymbolE10
  <12->   MnSymbolE12
}{}
\DeclareFontShape{OMX}{MnSymbolE}{b}{n}{
<-6>  MnSymbolE-Bold5
   <6-7>  MnSymbolE-Bold6
   <7-8>  MnSymbolE-Bold7
   <8-9>  MnSymbolE-Bold8
   <9-10> MnSymbolE-Bold9
  <10-12> MnSymbolE-Bold10
  <12->   MnSymbolE-Bold12
}{}

\let\llangle\@undefined
\let\rrangle\@undefined
\DeclareMathDelimiter{\llangle}{\mathopen}%
 {MnLargeSymbols}{'164}{MnLargeSymbols}{'164}
\DeclareMathDelimiter{\rrangle}{\mathclose}%
 {MnLargeSymbols}{'171}{MnLargeSymbols}{'171}
\makeatother

\catcode`,\active

\catcode`\,12

\def\XXint#1#2#3{{\setbox0=\hbox{$#1{#2#3}{\int}$}
     \vcenter{\hbox{$#2#3$}}\kern-.5\wd0}}

\renewcommand{\l}{\lambda}
\newcommand{\gym}{g_{\scalebox{0.45}{\text{YM}}}}

\newcommand{\Z}{\mathbb{Z}}
\newcommand{\rZ}{{\rm Z}}

\newcommand{\rt}{{\rm T}}

\newcommand{\CP}{\mathbb{CP}}
\newcommand{\wh}{\widehat}
\newcommand{\rr}{{\rm r}}

\DeclareMathOperator{\Tr}{Tr}
\DeclareMathOperator{\vol}{vol}

\def \del{\partial}

\def\ov{\over}
\def \ci {\cite}

\def \foot {\footnote}

\def\la{\label}
\def\foot{\footnote}
\newcommand{\rf}[1]{(\ref{#1})}
\def \OO {{\cal  O}}\def \no {\nonumber}

\def \ed {\small
\bibliography{BT-Biblio}
\bibliographystyle{JHEP-v2.9}
\end{document}}
\def  \iffa  {\iffalse}

\def \adsz  {AdS$_{4}\times S^{7}/\mathbb{Z}_{k}$}

\def \adssf  {AdS$_{7}\times S^{4}$}

\def \te {\textstyle}

\def\b{\beta}
\def\g{\gamma}

\def\k{\kappa}
\def\l{\lambda}

\def\s{\sigma}

\def\x{\xi}

\def \vp {\varphi}

\begin{document}

\begin{titlepage}
\begin{tabbing}
%\today \date{\today}
\hspace*{11.5cm} \=  \kill % set the tabbings
\>  PUPT-2648\\
\>  Imperial-TP-AT-2023-05 \\
\> %none
\end{tabbing}

%\today
\vspace*{15mm}
\begin{center}

\centerline{\large\sc
(2,0)   theory on $S^5 \times S^1$  and quantum  M2 branes 
%in twisted AdS$_7 \times S^4$ 
 % or  Quantum  M2 brane in twisted AdS$_7 \times S^4$  and superconformal index of (2,0)  theory
   }
   \vskip 4pt
%\centerline{\large\sc }

%{\Large\sc   title  }\vskip 9pt
%{\Large\sc    subtitle }

\vspace*{10mm}

{\large M. Beccaria${}^{\,a}$, S. Giombi$^{\,b}$, A.A. Tseytlin$^{\,c, }$\footnote{Also at   the Institute for Theoretical and Mathematical Physics (ITMP) of MSU    and Lebedev Institute.}} 

\vspace*{4mm}
	
${}^a$ Universit\`a del Salento, Dipartimento di Matematica e Fisica \textit{Ennio De Giorgi},\\ 
		and I.N.F.N. - sezione di Lecce, Via Arnesano, I-73100 Lecce, Italy
			\vskip 0.1cm
${}^b$  Department of Physics, Princeton University, Princeton, NJ 08544, USA
			\vskip 0.1cm
${}^c$ Blackett Laboratory, Imperial College London SW7 2AZ, U.K.
			\vskip 0.1cm
\vskip 0.2cm {\small E-mail: \texttt{matteo.beccaria@le.infn.it},\ \texttt{sgiombi@princeton.edu}, \ \texttt{tseytlin@imperial.ac.uk}}
\vspace*{0.8cm}
\end{center}

\begin{abstract}  
 The superconformal index  $Z$ of the 6d (2,0) theory on $S^5 \times S^1$
  (which is related to the   localization partition function  of 5d SYM on $S^5$) 
  should be captured at large $N$ by the quantum M2 brane theory in the dual M-theory background.
Generalizing the type IIA string theory limit  of this  relation discussed  in arXiv:2111.15493 and arXiv:2304.12340,  we 
consider  semiclassically quantized M2 branes  in a half-supersymmetric 11d background  which is a  twisted product of 
 thermal AdS$_7$  and $S^4$.  We show that  the  leading  non-perturbative term at large $N$   
 is reproduced precisely by the 1-loop  partition function of  an  ``instanton''   M2  brane 
  wrapped  on  $S^1\times S^2$ with  $S^2\subset  S^4$.  Similarly, the (2,0) theory   analog of the BPS  
  Wilson loop expectation   value  is reproduced  by the partition function of  a  ``defect''  M2 brane  wrapped  on  thermal 
  AdS$_3\subset$  AdS$_7$.  We  comment on  a  curious  analogy  of these results   with  similar 
 computations  in   arXiv:2303.15207 and  arXiv:2307.14112 
  of the partition function of quantum  M2  branes  in AdS$_4 \times S^7/\mathbb Z_k$ which 
  reproduced the corresponding  localization  expressions  in the    ABJM   3d gauge   theory. 
   \end{abstract}

\iffa 
 Superconformal index  $Z$ of 6d  $U(N)$    (2,0)    theory on $S^5 \times S^1$
  (that  may be interpreted as  localization partition function  of 5d SYM on $S^5$) 
  may be   expected to  be related  to M2  brane  partition   function in dual M-theory background. 
Generalizing   type IIA string theory limit  of this  relation   discussed  in arXiv:2111.15493 and  arXiv:2304.12340 
   we  consider quantum  M2 brane  in half-maximally supersymmetric  11d   background  which is a ``twisted version 
 AdS$_{7,\b} \times \tilde S^4$. Here  AdS$_{7,\b}$   has  boundary  $S^5 \times S^1_\b $
 ($\b$ is the length of the circle coordinate $y$)  and ``twisted''  4-sphere  $ \tilde S^4$  has one   angle shifted  by $iy$. 
 %The dual theory   (2,0)  6d  theory  on $S^5 \times S^1$  with an R-symmetry   twist. 
We show that  the  leading  large $N$  non-perturbative term 
 $(4\sinh^{2}\frac{\beta}{2})^{-1}\,e^{-\beta N}$   in  $Z$ 
 is reproduced precisely by the 1-loop semiclassical partition function of  an  ``instanton''   M2  brane 
  wrapped  on  $S^{1}_{\beta}\times S^{2}$ with  $S^{2}\subset \tilde S^{4}$. 
 Similarly, the (2,0) theory   analog of   BPS  Wilson loop expectation   value $(2\sinh\frac{\beta}{2})^{-1}\, e^{N\beta}$
  is reproduced  by the partition function of  a  ``defect''  M2 brane  wrapped  on  AdS$_{3,\b}\subset$  AdS$_{7}$.
 We  comment on  a  curious  analogy  of these results   with  similar 
 computations  in   arXiv:2303.15207 and  arXiv:2307.14112 
  of the partition function of quantum  M2  brane  in AdS$_4 \times S^7/\mathbb Z_k$ which 
  reproduced the corresponding  localization  expressions  in the   % $U_{k}(N) \times U_{-k}(N)$ 
  ABJM   3d gauge   theory. 
\fi 
	%Keywords: {\sc insert here keywords}
	
	\date{today}
\end{titlepage}
%\end{comment}
%%%%%%%%%%%%%%%%%%%%%%%%%%%%%%%%%%%%%%%%%%%%%%%%%%%%%%%%%
%%%%%%%%%%%%%%%%%%%%%%%%%%%%%%%%%%%%%%%%%%%%%%%%%%%%%%%%%

\tableofcontents
\vspace{1cm}

\def \T {{\rm T}}

\def \RR {\mathbb R} \def \ZZ {\mathbb Z}
\def \four  {\tfrac{1}{4}}
\def \AdS {{\rm AdS}}
  \def \tS  {\tilde S}
\def \wZ  {{\wh Z}} 
\def \V  {{\rm V}}  \def \G {\Gamma} \def \Ze {{\cal Z}} 
\def \ha {{1\ov 2}}  \def \nb {n_{_\b}}
 \def \rZ  {{\rm Z}}
\def \Ze {{\cal Z}}
\def \A  {{\cal A}} 
\def \y  {{\rm y}}  \def \rb  {{\rm b}}

\newpage
\setcounter{footnote}{0}
\section{Introduction}

The 6d  (2,0)   superconformal field theory should be describing the low-energy dynamics of $N$ coincident M5 branes.  
It is expected to be dual  \ci{Maldacena:1997re,Aharony:1999ti}  to 11d  M-theory  %or quantum supermembrane (M2  brane) 
theory  on the \adssf\  background, 
which is a limit of  the M5 brane solution of 11d supergravity \ci{Gueven:1992hh,Gibbons:1993sv}\foot{Here $ds^2_{\AdS_{7}}$  and $ds^2_{S^{4}}$  are the metrics of the unit-radius 
AdS$_7$ and $S^4$.  We shall often use the notation $dS_n \equiv  ds^2_{S^{n}}$. 
  $\ell_P$ is the 11d Planck constant  related to the   gravitational constant  in the (Euclidean) 11d   supergravity action 
  $S_{11} =  -{1\ov 2 \k_{11}^2} \int d^{11} x \sqrt G ( R - {1\ov 2 \cdot 4!  } F_{MNKL}^2 + ...) $ as 
   $2 \kappa^2_{11}= (2 \pi)^8 \ell_P^9$   
    and to  
  the  M2  brane tension  as  $T_2= {1\ov (2\pi)^2 \ell_P^3}$.
   Also, $\vol_{S^4}$ is the normalized  volume 4-form of $S^4$, i.e. 
     $\int_{S^{4}}\vol_{S^4}=1$ with $  \vol(S^{4})=\frac{8\pi^{2}}{3}$.}
\ba
\la{1}
ds^{2}_{11}  = a^{2}\,\Big(ds^2_{\AdS_{7}}+{\four } ds^2_{S^{4}}\Big), \qquad 
\qquad  F_{4} &= dC_{3} = \pi^{2}a^{3}\vol_{S^4}\ , \qquad a^{3}=8\pi\, N\, \ell_{P}^{3} \ . 
\ea 
Due to the lack of an intrinsic  definition of the (2,0) theory and having only $N$  as a free parameter, 
it is  not clear how to define non-trivial observables (computable, e.g., by localization) 
that can be used to test  this AdS/CFT duality.\foot{Almost all  of the 
available information comes   from the  11d supergravity  effective action   and supersymmetry considerations 
 that  may  be used, e.g.,  to determine  the  M-theory  predictions for the 
a- and c- conformal anomaly coefficients of the (2,0)  theory (see, e.g.,  \ci{Henningson:1998gx,Tseytlin:2000sf,Beccaria:2014qea,Beem:2014kka,Ohmori:2014kda})
%v2
and thus, in particular, the expression for  its free energy  on $S^6$ 
(that should have the same structure as the free energy of the $\N=4$ SYM on $S^4$): $F \sim {\rm a}(N) \log \Lambda + $const. 
%,Beccaria:2015ypa}). 
One may also find a defect conformal anomaly by using M2  brane probe in \adssf\ background   as   discussed in \ci{Drukker:2020dcz,Drukker:2020swu}
and refs. there.}
%%%%%%%%%%%%%%%%%%%%%%%%%%
\iffa
\cite{Beccaria:2015ypa}  for a review).
In particular, one finds that the two  conformal anomaly  coefficients are given by 
$a=  -\frac{1}{288}\, (N-1)\big[( 2N+1)^2 + { 3\ov 4}\big] $, \ 
$c =    -\frac{1}{288}\, (N-1) ( 2N+1)^2$.
The leading  $N^3$ terms   follow    \ci{Henningson:1998gx}
from the classical supergravity  action,  
the   order $N$ terms   originate from the  $R^4$   corrections to the  11d supergravity action  \ci{Tseytlin:2000sf}
and order $N^0$ terms  --  from  the 1-loop 11d  supergravity 
corrections \ci{Beccaria:2014qea,Mansfield:2003bg}. 
The exact expressions follow also  from  non-perturbative approaches   based on 
supersymmetry  constraints   \cite{Beem:2014kka,Ohmori:2014kda}.
\fi

To introduce an extra parameter one  may consider some  ``orbifolding''
 of \rf{1} (by analogy, e.g.,  with the ABJM theory \ci{Aharony:2008ug}  of multiple M2  branes  on $\RR^8/\ZZ_k$  dual to M-theory on \adsz).
 One option is to consider the (2,0)  theory on    $S^5 \times S^1_\b$    where 
 $\beta$  is the length of the circle. The dual M-theory  background may  then have the $\AdS_7$  part  with the 
 corresponding  $S^5 \times S^1_\b$  boundary, i.e. %  case  we will have 
 $ds^{2}_{\AdS_{7}}= dx^{2}+\sinh^{2}x\, dS_{5} + \cosh^{2}x\,dy^{2}$ where   $y\equiv y + \b$
 and $dS_5 $ is the  metric of a unit-radius 5-sphere.\foot{In general,    introducing a thermal 
 circle one would  need  to  consider also
 %M1
  black hole like   geometry  with the corresponding asymptotics \cite{Witten:1998zw}.  
  This   will not be the case here  as  we will be interested in the  background  corresponding to a superconformal index with an extra R-symmetry twisting and periodic fermions.}

 Dimensionally reducing on  the $y$-circle, i.e. considering the limit of $\b\to 0$,  the M5 brane solution will reduce to the D4 brane solution of type IIA 10d supergravity, while  the (2,0) theory  on $S^5 \times S^1_\b$  is expected to be related to 
 the maximally supersymmetric  
 5d SYM theory on $S^5$. The 5d 
 SYM  theory does not have a %A15 
 first-principles    %non-
% perturbative
  definition  
  being  nonrenormalizable, i.e.  
  the  (2,0)   theory  should be thought of as its UV completion
   (cf. \cite{%Seiberg:1996bd,
   Seiberg:1997ax,Douglas:2010iu,Lambert:2010iw}).
  Yet  this relation  may be useful at 
   a heuristic level  as  one may   attempt to define   free energy of the SYM theory on $S^5$ 
  by analogy with 4d  SYM  theory where it can be computed  from localization. 
 
 It turns out that the requirement of preservation of 16 real supersymmetries  demands introducing 
 an extra  R-symmetry twist in the (2,0)  theory on $S^5 \times S^1_\b$, or a twist in the $S^4$ part of the background \rf{1}.
 This was understood in \ci{Bobev:2018ugk}   when constructing the   type IIA solution  which  corresponds to a
   D4  brane  
   %A19
   world volume 
   wrapped on $S^5$.
    The  11d  uplift of this  solution  is related    by an 
 analytic continuation to the   following 11d background 
 %v3
 \ci{Bobev:2018ugk,Mezei:2018url,Gautason:2021vfc}
 \ba
\la{2}
ds^{2}_{11} &= a^{2}\Big(\big[dx^{2}+\sinh^{2}x\, dS_{5} +\cosh^{2}x\,dy^{2}\big] +\four\big[
du^{2}+\cos^{2}u\,dS_{2}+\sin^{2}u\, (dz+ idy)^{2}\big]\Big),  \\
%v4
C_{3} &=- \tfrac{1}{8}a^{3}\,\cos^{3}u\, \vol_{{S}^{2}} \wedge (dz+idy)\ . \la{3}
\ea
Here  the $S^4$ part  $du^{2}+\cos^{2}u\,dS_{2}+\sin^{2}u\, dz^{2}$  got the $2\pi$ periodic angle  $z$  shifted by $iy$  where 
$y\in (0, \beta)$ is the circular  11d coordinate.\foot{This    complex  background   becomes real  after $y\to i t$
with   the   time-like  direction $t$ here playing the role of the 11d circle. 
%v4
Note that here $u\in (0, {\pi\ov2})$ and $z\in (0, 2 \pi)$.}
This  background is 
related  to \rf{1}  by a periodic  identification and a   coordinate shift 
so is an obvious solution  of the  11d   supergravity.\foot{Note that  near $x=0, \ u=0$ and  relevant part of the metric 
becomes $dy^2 + du^2 + u^2 (dz + i dy)^2$  so that it may be   thought of as  a  special case of a (complex or time-like) 
Melvin twist  discussed in    \ci{Gibbons:1986cq,Russo:1995ik,Tseytlin:1995zv}  and,  in particular, 
 in 11d context  in \ci{Russo:1998xv}).
}
We will denote  the  first 7d    part   of \rf{2}   as $\AdS_{7,\b}$   and the 4d part as $\tS^4$
%A15
and  somewhat loosely refer to \rf{2} as a ``direct product'' $\AdS_{7,\b}\times \tS^4$.

 Our aim  in this paper is to consider the quantum M2  brane in the $(N,\beta)$  dependent  background \rf{2},\rf{3} 
 and compute its partition  function  in the semiclassical  (large tension $\T_2=  a^3 T_2 = {2\ov \pi} N\gg1 $) expansion near particular  classical solutions  %A15with non-degenerate  3-volume, 
 similarly to   how that was done  in the \adsz\ case in \ci{Giombi:2023vzu,Beccaria:2023ujc}. 
 This will  represent   an M-theory   %A15 (fixed $\beta$) 
 generalization of the type IIA string theory semiclassical 
  computations  done  in the limit  $\beta\to 0, \ N\to \infty$ with fixed  $N \beta$ 
   in \ci{Gautason:2021vfc,Gautason:2023igo}.

 We will provide a  check  of the  AdS$_7$/CFT$_6$   correspondence   in this setting 
  by establishing   matching of  quantum  M2  brane results  with 
 % particular  terms  in 
  the  large $N$ expansion of the 
   %v2
  supersymmetric  partition function of the 
 (2,0) theory on $S^5 \times S^1_\b$   with R-symmetry twist  (and  periodic fermions), 
 identified  with the corresponding superconformal index computed in  \cite{Kim:2012ava,Kim:2012qf}.\foot{Ref. \cite{Kim:2012ava} 
   started with  the abelian 6d $(2,0)$ theory (i.e. tensor multiplet)  with 32 supersymmetries and by introducing 
a Scherk-Schwarz-like  R-symmetry twist 
  obtained a theory on $S^{5}\times S^{1}$ with 16 supersymmetries and a subgroup $SO(2)\times SO(3)$ of the 
original $SO(5)$ R-symmetry. 
The  $SO(2)\subset SO(5)$ twist was necessary to   have  constant spinors on $S^{5}$.
Upon dimensional reduction,
%AA16 
 %in the 5d SYM theory 
the R-symmetry twist  leads to  extra mass terms  in the 5d  SYM action.
The construction was then 
 extended to  the non-abelian case via 5d SYM connection, 
   and  using supersymmetric localization  provided the  expression for the 
  perturbative 
partition function in the  form of a matrix model  \cite{Kim:2012ava},  which was   supplemented by all  
instanton corrections  in \cite{Kim:2012qf}.
%The matrix model obtained by this  construction does not compute the partition function of the $(2,0)$ theory.
 The $SO(2)$ twist  corresponds  to the  introduction of 
% inserts in the standard partition function
  a chemical potential coupled to the  R-charge and  the  corresponding localization 
   matrix model  computes 
the (unrefined) superconformal index of the $(2,0)$ theory (see also \ci{Kim:2016usy}).
%M1
}

%v2 
As the non-abelian (2,0)  theory does not have an explicit Lagrangian formulation,   its supersymmetric partition function  on $S^5 \times S^1$  that should be equal  to the index cannot be   computed directly, but it may be interpreted as a 
%localization result for 
 partition function of the 5d SYM  theory  (assuming the latter has a well-defined 
 UV completion).
 % (the localization expression  is determined by an integral of the 1-loop determinants  times instanton contr 
Then  the  superconformal  index  may   be interpreted as  a (properly defined)  localization result for the 
%supersymmetric 
   partition function of  5d SYM on $S^5$   with $\gym^2$  proportional to $\b$   up to a length scale  factor. 
    By analogy with the  4d SYM theory, 
    this   suggests also  to consider the localization %or index-related 
   expression for  the BPS Wilson loop expectation value (cf.  \cite{Young:2011aa}) 
    which may be then compared 
   with an M2  brane  semiclassical computation as in  \ci{Giombi:2023vzu}. 
   
   %%%%%%%%%%%%%%%%%%%%
   \iffa 
   %v2
   add more comments?  like 
   
   In this paper, we will consider the supersymmetric partition function of the N = (2, 0) theory itself on S 1 � S 5 , which is closely related to the 6d superconformal index. Although this partition function cannot be computed directly in six dimensions, due to the absence of a useful lagrangian formulation, it has been computed recently using maximally supersymmetric Yang-Mills theory in five dimensions on S 5 [9?11] 1 . It relies on the conjecture that the nonperturbative physics of 5d SYM allows us to extract non-trivial dynamics of the 6d (2,0) theory with circle compactification [17, 18]. The most important part of the dictionary is that the five-dimensional gauge coupling g 2 is related to the ...................... so that strong coupling corresponds to large radius. Although this five-dimensional theory is non-renormalizable, it is conjectured that by including non-perturbative contributions in five dimensions, one can capture all of the protected states contributing to the 6d superconformal index. This is remarkable given that the 5d partition function is computed as an instanton expansion in powers of e ?4? 2/?, while the 6d superconformal index is naturally an expansion at large radius in powers of e ?? with integer coefficients.

I would add to this  that  computation of localization free  energy in 5d SYM is 
a  conjecture in itself -- it assumes  that  Z is again   given by an integral of 1-loop expression that is well-defined despite 
the theory being non-renormalizable -- i.e. one assumes that localization argument 
applies despite path integral not really being well defined (if one assumes  that UV cutoff is fixed one is to be sure that 
it is consistent with susy etc). It works  somehow  beacuse   of maximal susy  what is effectively counted-- as in index-- are only bps  states...  This is not true in general for other partition functions computed from localization. 
   
   \fi 
   %%%%%%%%%%%%%%%%%%%%%%%%%%

Denoting  by $Z_{N}(\beta)$ the index of the $(2,0)$ theory on $S^{5}\times S^{1}_{\beta}$
one  finds for the  large $N$, fixed $\beta$ expansion  of the  corresponding free  energy 
 $F_{N}(\beta) = -\log Z_{N}(\beta)$     \cite{Kim:2012ava,Kim:2012qf}
 \ba
\la{4}
&F_{N}(\beta) = F_{N}^{\rm pert}(\beta)+F_{N}^{\rm np}(\beta)\ , \qquad \qquad 
F_{N}^{\rm pert}(\beta)= -  (\tfrac{1}{6} N^3 - \tfrac{1}{8} N)\, \b  + \sum_{n=1}^\infty  c_n\,  e^{- n \beta}
\ , \\
 & F_{N}^{\rm np}(\beta) = \frac{1}{4\sinh^{2}\frac{\beta}{2}}e^{-N\beta}+\mc O(e^{-2N\beta})  \ . \la{5}
\ea
%%%%%%%%%%%%%%%%%%%%%%
For the  natural analog of  the Wilson loop  one finds
 for large $N$ 
     \cite{Kim:2012ava,Kim:2012qf}
% One can also consider in the 5d SYM theory a supersymmetric Wilson loop
%$W$ compatible with the supercharge used in localization. For its expectation value, localization predicts at large $N$
\be
\la{6}
\langle W \rangle = \frac{1}{2\sinh\frac{\beta}{2}}e^{N\beta} +  \OO(N^0) \ . 
\ee
Below   we will  reproduce the expressions (\ref{5})  and (\ref{6})  on the M-theory side,  by  performing semiclassical 
M2-brane  computations in the background \rf{2},\rf{3}. 
In   the  case of the  non-perturbative contribution to free energy  in \rf{5}, 
the classical  M2    brane  solution will   be wrapped on 
  $S^{1}_{\beta}\times S^{2}$ with $S^{1}_\b \subset \AdS_{7,\beta}$  and 
$S^{2}\subset \tilde S^{4}$. In the case of the Wilson  loop  \rf{6}, 
the dual  M2 brane solution will be wrapped on  AdS$_{3,\beta}\subset \AdS_{7,\beta}$, where 
AdS$_{3,\beta}$ is the ``thermal'' AdS$_3$ background. 

In both cases, the exponents   in \rf{5} and \rf{6}   will come from the classical M2  brane action 
while the  $\beta$ dependent prefactors will be precisely reproduced by the one-loop M2  brane fluctuation determinants 
as in  \ci{Giombi:2023vzu,Beccaria:2023ujc}.
Our  results   will generalize to the finite $\b$ case 
the   analogous   computations in the type IIA string-theory limit in  \cite{Gautason:2021vfc,Gautason:2023igo}.
% They will also be technically more straightforward and analytic (we will not need  to use numerical evaluation). 

%A15
The plan of the paper is as  follows. 
In section \ref{sec:loc} we review the localization results for the $(2,0)$ theory  superconformal 
 index and the  analog of the supersymmetric Wilson loop,  leading to (\ref{5}),(\ref{6}). 
In section \ref{sec:setup} we  discuss the    general    structure of the  M2  brane semiclassical partition function.
Section \ref{sec:calc-free} presents the details of the   calculation of this   partition  function 
in the case of the  $S^1_\b \times S^1$  M2 brane  instanton    background  reproducing (\ref{5}). 
%After mode expansion 
% along $S^{1}_{\beta}$ the action is reduced to a sum of terms for each fluctuation which is a massive scalar in  $S^{2}$.
 Section \ref{sec:calc-wl}  addresses   similar computation in the  case of the M2   brane 
 wrapped  on  $\AdS_{3,\beta}$   reproducing the Wilson loop expectation value in \rf{6}. 
%analyze the M2 brane relevant for the supersymmetric Wilson loop, now . Technically, the analysis is different
%since the mode expansion along $S^{1}_{\beta}$ would introduce position dependent mass terms. In this case, it is convenient not to expand in modes 
%and exploit instead results for functional determinants in thermal $AdS_{3}$, suitably generalized for our purposes. 
Section  \ref{sec-concl}  contains  a summary and concluding remarks.
  Appendices contain    some     technical details   used  in the main  part of the paper.
 %complementary material and discussions.

%%%%%%%%%%%%%%%%%%%%%%%%%%%%%%%%%%%
\section{Localization  expressions  for the  free  energy  and Wilson loop }
\la{sec:loc}

The superconformal index 
of  $U(N)$   (2,0)   theory on $S^5 \times S^1_\beta$ was  found 
 \cite{Kim:2012ava,Kim:2012qf}  to be given by a matrix model 
 which is  the same as for the supersymmetric  3d  pure
 Chern-Simons theory  solved in \cite{Marino:2004eq}.
The result 
 may be represented as a product of two factors\footnote{Here 
the Dedekind function is $\eta(\tau) = q^{\frac{1}{24}}\prod_{n=1}^{\infty}(1-q^{n})$ where $q=e^{2\pi i \tau}$. Its modular 
transformation is $\eta(-1/\tau) = \sqrt{-i\tau} \ \eta(\tau)$.
}
 \ba \la{2.1}
& Z_N(q) \equiv e^{- F_N(\b)}=  Z^{(0)} _N(q)\,  Z^{\rm inst} _N(q)\ , \qquad \ \    q\equiv e^{-\beta} \ , \\
&Z^{(0)} _N(q) = \Big(\frac{\beta}{2\pi}\Big)^{N/2}e^{\frac{N(N^{2}-1)}{6}\beta}\prod_{n=1}^{N-1}(1-e^{-n\beta})^{N-n} \ ,
 \qquad 
Z^{\rm inst} _N(q) = \Big[\eta\Big(\frac{2\pi i}{\beta}\Big)\Big]^{-N} \ . \la{22} 
\ea
%e^{- F_N(\b)} \ , \qquad  F_{N}(\beta) = F_{N}^{(0)}(\beta)+F_{N}^{\rm inst}(\beta),\qquad 
%q\equiv e^{-\b}  \ . \la{21} \ee
 We shall  refer to $Z_N$ as  partition function. $F_N(\b)=-\log Z_N(q)$ may be  interpreted as a 
``supersymmetric'' free energy.\foot{In the interpretation 
of  $F_N(\b)$   as a     free energy of 5d SYM   theory on $S^5$   one may set  $
\beta = \frac{\gym^{2}}{2\pi R}$ where $R$ is an effective  length scale.}
%(cf. also {Minahan:2016xwk}). } 
%Explicitly, one finds\ba\la{22}
%F_{N}^{(0)}(\beta) =& -\log\Big[\Big(\frac{\beta}{2\pi}\Big)^{N/2}e^{\frac{N(N^{2}-1)}{6}\beta}\prod_{n=1}^{N-1}(1-e^{-n\beta})^{N-n}\Big], \qquad
%F_{N}^{\rm inst}(\beta) = N\,\log\eta\Big(\frac{2\pi i}{\beta}\Big),
%\ea
%where the second term   originates 
 % from the instanton parts of the matrix model partition function.
 % while the first term is what remains once instantons are neglected.
%\footnote{It is convenient to avoid the standard terminology that calls $F_{N}^{(0)}$ the ``perturbative'' part, since in our context the splitting 
%into  perturbative and non-perturbative contributions refers to the absence of presence of $\exp(-N\beta)$ corrections.}

To study the expansion of the partition function $Z_N$ 
at large $N$ and fixed  $\beta$, it is convenient to apply a modular transformation to the $\eta$-function
 factor $Z^{\rm inst} _N(q)$  in  $Z_N$. This     gives  % after a short calculation
\ba
\la{2.4}
Z_{N}(q) =&   \ 
 q^{\epsilon_{0}(N)}\, \wh Z_{N} (q) \ , \qquad \ \ \ \ \ \ 
 \epsilon_{0}(N) = -\tfrac{1}{6} N(N^{2}-1)-\tfrac{1}{24} N= -\tfrac{1}{6} N^3 + \tfrac{1}{8} N   \ , \\
 \wh Z_{N} (q) =&\prod_{n=1}^{N}\prod_{m=0}^{\infty}\frac{1}{1-q^{n+m}}\ , \la{24}
\ea
 where $\epsilon_{0}(N)$ is the  ``supersymmetric Casimir energy'' 
 %v2
 \cite{Kim:2013nva,Assel:2015nca,Bobev:2015kza,BenettiGenolini:2016qwm}.
 
   The partition function \rf{24} 
   has  an expansion in powers of $q$ with integer $N$-dependent coefficients.
    The coefficients take finite values for large $N$: the  $N\to\infty$ limit  $\wh Z_{\infty}(q)$ 
 of $\wh Z_N$ is   the MacMahon function   \be
\la{2.6}
\wh Z_{N}(q) \ \stackrel{N\to \infty}{\to} \  \wh Z_{\infty}(q) \ , \quad \qquad 
\wh Z_{\infty}(q) =  \prod_{n=1}^{\infty}(1-q^{n})^{-n} = 1+q+3q^{2}+6q^{3}+13q^{4}+24 q^{5}+\cdots\, .
\ee
The expression  $\wh Z_{\infty}$  may be interpreted as 
%M1
the unrefined
superconformal  index counting  BPS   states  of 11d  
supergravity on  $\AdS_{7}\times S^{4}$, i.e. 
%A16
given  by the   sum over  Kaluza-Klein states of the $S^4$  compactification  \cite{Kim:2012ava}.

Finite $N$ corrections to the partition function 
can be read off from (\ref{24}) after writing it in the following 
equivalent form
%A19
\be
 \wh Z_{N}(q) = \wZ_{\infty}(q) \prod_{n=0}^{\infty}\prod_{m=0}^{\infty}(1-q^{N+n+m+1}) \ . \la{26}
\ee
Expanding $\log\wh Z_{N}$ in powers of $q^{N}$, summing over $n,m$, and exponentiating back   gives
%\foot{This   may be interpreted as  a ``giant graviton''
% expansion \cite{Arai:2020uwd,Gaiotto:2021xce}.}
 \ba
\la{2.8}
\wh Z_{N}(q) &= \wh Z_{\infty}(q) \Big[1-\frac{q}{(1-q)^{2}}\,q^{N}+\frac{2q^{3}}{(1-q^{2})^{2}(1-q)^{2}}\,q^{2N}+\cdots\Big] \lp
= \wh Z_{\infty}(q)\Big[1-\frac{1}{4\sinh^{2}\frac{\beta}{2}}e^{-N\beta}+\frac{1}{32\,\sinh^{4}\frac{\beta}{2}\cosh^{2}\frac{\beta}{2}}\,e^{-2N\beta}+\cdots \Big]\ . 
\ea
Combining (\ref{2.4}) and (\ref{2.8}), we can write the large $N$,  fixed $\beta$  expansion 
 of the  free energy $F_N$  in (\ref{2.1}) as 
a sum of a perturbative   and non-perturbative parts
\ba
\la{2.9}
F_{N}(\beta) &= F_{N}^{\rm pert}(\beta)+F_{N}^{\rm np }(\beta) \ , \\
%\qquad 
F_{N}^{\rm pert}(\beta) &=\epsilon_{0}(N) \,\beta +  \wh F(\b) 
% f_{_{\rm BPS}}(\beta)
 \ ,  \qquad \ \ \ \   %f_{_{\rm BPS}}
\wh F (\b)\equiv
 - \log \wh Z_{\infty} (q)= \sum^\infty_{n=1} c_n\,  e^{-n \beta}\  , \la{29}\\
\la{2.10}
F_{N}^{\rm np}(\beta) &=  \frac{1}{4 \sinh^2\frac{\beta}{2}}\,e^{-N\beta}+\mc O(e^{-2N\beta}) \ , 
\ea
%where the order $N^0$ term $f_{\rm KK-BPS}$ is the log of the MacMahon function. 
where $\epsilon_{0}(N)$ is   given in \rf{2.4}   and  $c_n$ in \rf{29}  following from \rf{2.6}
are  $c_1=-1, \ c_2 = - {5\ov 2},\ c_3 = - {10\ov 3},...$.

%AT15
The leading   $N^3 \b$ term in the perturbative part of   (\ref{29})  where
%v2
 $\eps_0= - {1\ov 24} (4N^3 - {3} N) $ as  in \rf{2.4} 
  should   originate from  the  %be compared with the 
11d supergravity  action  $\int R + ...$  evaluated on the  corresponding dual 
background  $ \AdS_{7,\b }\times \tS^{4}$ in  \rf{2},\rf{3}.\footnote{%v2   
The computation of the $N^{3}$ term in the free energy 
 from the  supergravity action in  thermal $\AdS_{7}\times S^{4}$  has a priori   no reason to  match 
 the  coefficient in the index asymptotics,  see a   discussion  in Appendix \ref{app:vol}.
Reproducing  the  coefficient of this   leading $N^{3}$ contribution 
  attempted  in \cite{BenettiGenolini:2016tsn,Bobev:2019bvq}  requires adding 
finite ``counterterms''  to the  low-dimensional effective 
supergravity action  that were claimed to be needed to preserve supersymmetry. 
%v2
%This claim was further  corroborated in an   unpublished work of P. Bomans where (2,0)  theory was %considered on 
% more general ${SE}^5 \times S^1$  backgrounds.
Let us also note  that, in view of the relation  between  the supersymmetric  Casimir energy and the c-coefficient of the conformal anomaly \ci{Bobev:2015kza},
 %(see below), 
 one may expect  that to match the former on the supergravity side one may need a more subtle  procedure than just directly  evaluating the supergravity action on the $\AdS_{7}\times S^{4}$   background:
 to  capture the c-anomaly one needs to perturb the $\AdS_7$ boundary metric to have a non-zero 6d Weyl tensor 
  \ci{Henningson:1998gx}. } 
The first  subleading  $ N\b$   term in $F_{N}^{\rm pert}$  
%v2 
should 
% is expected to
   originate from  the $R^4$ invariant in the 11d  effective  action, 
by   analogy with   
    the case of the 11d effective action evaluated on the standard 
AdS$_7 \times S^4$  background, reproducing  \ci{Tseytlin:2000sf}  
 the  order $N$ term in the  coefficient 
a$= 4 N^3 - {9\ov 4} N - {7\ov 4}$ 
of the conformal anomaly  of  the (2,0) theory on $S^6$.\foot{Note that while  in  the case  of AdS$_7$ with  $S^6$   boundary   the   value of 11d effective action 
  is proportional to $\vol(\AdS_7) = { \pi^3 \ov 3} \log \eps$
  %A16
    (where $\eps\to 0$ is an IR cutoff)  and thus    computes the a-anomaly coefficient, in the case of 
  the $S^5 \times S^1_\b$ boundary we have    
  $ \vol(\AdS_{7,\b}) = -\frac{5\pi^{4}}{48}\b$  (see Appendix \ref{app:vol})   and thus the  local $\int (R +R^4)$  part of the  11d 
  effective  action evaluated on AdS$_{7,\b} \times S^4$  is finite and  linear in 
   $\b$. 
  }  
  %v2
  Let us note that in general the supersymmetric Casimir   energy of a  6d (2,0)  supersymmetric   theory on $S^5 \times S^1$ 
  should be related to the   conformal c-anomaly   coefficient as \ci{Bobev:2015kza}  \ $\epsilon_0 = - {1\ov 24}   {\rm c}$. 
  For the $SU(N)$  (2,0)  theory   one has ${\rm c}= 4 N^3 - 3N -1$
    \ci{Henningson:1998gx,Beem:2014kka}
  which is thus consistent with \rf{2.4}  (the $-1$ term is absent in the $U(N)$ case).

%A16
The  term $\wh F(\b) $  in  \rf{29}  should be 
%v3
 reproduced  by the 1-loop 11d supergravity partition function on 
 $ \AdS_{7,\b }\times \tS^{4}$ (with periodic boundary conditions on fermions).
 %\foot{
 The  supergravity index  $\wh Z_{\infty}(q)$  was found  in \cite{Kim:2012ava}  from  the BPS   KK  spectrum  of  $S^4$  
 compactification of 11d supergravity \ci{Gunaydin:1984wc,Bhattacharya:2008zy}, 
  adding also an  R-charge  shift to the  Hamiltonian (conjugate to the Euclidean ``time'' $y$)
    when  defining  the index. This
    %A16
      R-charge shift   corresponds  effectively to computing a supersymmetric partition  function on 
 $ \AdS_{7,\b }\times \tS^{4}$   with the $z\to z + i y$ shift in a $S^4$ angle as in \rf{2}. 
 % counts   BPS states  for  standard  $S^4$ compactification 
  %and the distinction between $\tS^4$ and $S^4$  (due to coordinate shift) 
 % should not be visible at the level of 1-loop supergravity correction. 
There is again an analogy with how the constant $N^0$  term in the a-coefficient of 6d conformal anomaly
  is found from the 1-loop 11d supergravity effective action 
  on $ \AdS_{7}\times S^{4}$  with $S^6$ boundary \ci{Beccaria:2014qea}.
  %} 

%\footnote{The first principle determination of the counterterm coefficients 
%is  an open issue. How to determine the linear in $N$ part of the index Casimir term $\frac{N}{8}\beta$ is even more unclear.}

%A16
The  large $N$ expansion (\ref{2.8}) of  the superconformal index of the $(2, 0)$     theory  
   was  interpreted  in \cite{Arai:2020uwd,Imamura:2022aua}
   as representing  the 11d    supergravity index   $\wh Z_{\infty}(q)$
corrected by the   contributions of   other 
 BPS states  corresponding   to     wrapped    M2  branes
(that  here  play the role of ``giant gravitons'', cf. \cite{Gaiotto:2021xce}). %,Lee:2022vig}).
%\foot{In  \cite{Arai:2020uwd}
%the  terms in the bracket in \rf{2.8} were written as  
%$1+\sum_{n+m=1}^{\infty}\wh{\mathrm{I}}_{n,m}(q), \ \  \wh{\mathrm{I}}_{n,m}(q) = q^{(n+m)N}\big(1+\mc O(q)\big) .
%$
%   It was conjectured   that this  sum represents  the combinations of 
%  contributions   of two distinct supersymmetric M2-brane embeddings
%   wrapping $S^{1}_{\beta}\times S^{2}$ with $S^{2}\subset S^{4}$.}
%It was  also  noted that  $\wh{\mathrm{I}}_{n,m}(q)$  are the same as 
% superconformal indices of the ABJM theory living on the 
%stack of $n+m$ supersymmetric M2-branes. 
%At leading order $n+m=1$, \ie considering $(1,0)+(0,1)$, one has a single M2 brane hosting 
%a \underline{free theory} for which the index is easily computable.
%This gives the result obtained in \cite{Arai:2020uwd}. Extension to higher $(n,m)$ is far from trivial, see \cite{Imamura:2022aua}.
%A somewhat more general discussion of these mechanism is in \cite{Gaiotto:2021xce,Lee:2022vig}.

Below we   will  prove  that the leading $ -[4\sinh^{2}\frac{\beta}{2}]^{-1}  e^{-N \b}$  term in \rf{2.8} or in \rf{2.10} 
 originates precisely from  the partition function  of M2  brane  wrapped on $S^1_\b \times S^2$, 
 in full analogy with how that  happened  \cite{Beccaria:2023ujc}  for the instanton 
 M2  brane in \adsz\   background  in the ABJM  case. 
 %This will  generalize  the type IIA   str

%It will be captured by a semiclassical analysis around a certain wrapped M2 brane. The exponential factor will be 
%identified with the classical M2 volume, while the prefactor will come from the fluctuations one-loop functional determinant.

%\paragraph{Supersymmetric Wilson loop}
By analogy with the familiar $\mc N=4$ SYM case \ci{Drukker:2000rr}, it is possible to insert into the 
matrix model integral    found in \cite{Kim:2012ava,Kim:2012qf}
a counterpart of the  Wilson loop   operator $W(X) = \Tr e^{X}$ (where $X$ is the matrix   which is the integration variable). 
 One  may interpret $\langle W\rangle$ as the expectation value 
of a suitable \cite{Kim:2012ava,Kim:2012qf}
 supersymmetric Wilson loop in the  SYM theory on $S^{5}$  (cf. \cite{Young:2011aa})\foot{As 
 discussed  in \cite{Kim:2012qf}, representing $S^{5}$ as a Hopf fibration over $\mathbb{CP}^{2} $
 suggests  the following field theory analog  of this   operator:
$
W = \Tr\Big[\mc P\, \exp\oint ds\,\big(iA_{m}\dot x^{m}+\phi\, |\dot x|\big)\Big],$
%\ee
where  $x^{m}(s)$ wraps the Hopf fiber.} % located at a point  $\mathbb{CP}^{2}$ base.} 
or  rather of a  corresponding 2-defect  operator in the (2,0)   theory on $S^5 \times S^1_\b$ 
 that   wraps   $S^1$ of   $S^5$   as well  $S^1_\b$. 
%\foot{The 5d supersymmetric Wilson loop should respect the supercharge that has been used to localize the twisted 5d SYM theory in \cite{Kim:2012ava}.
%As shown in \cite{Kim:2012qf}, representing $S^{5}$ as a Hopf fibration over $\mathbb{CP}^{2} = S^{5}/U(1)$, see \eg \cite{Duff:1998us}, 
%leads to the following field theory identification of (\ref{2.11})
%\be
%W = \Tr\Big[\mc P\, \exp\Big(\oint ds\,(iA_{m}\dot x^{m}+\phi\, |\dot x|)\Big)\Big],
%\ee
%where the curve $x^{m}(s)$ wraps the Hopf fiber, located at a point on the $\mathbb{CP}^{2}$ base. Computing the expectation value 
The resulting  matrix model  expectation value is   \cite{Kim:2012ava,Kim:2012qf} 
(using  the  original  Wilson loop   computation in  $U(N)$  Chern-Simons matrix model  \ci{Witten:1988hf,Kapustin:2009kz}) 
\be
\la{2.13}
% N\gg 1: \qquad\qquad  
\langle W \rangle = e^{N\beta\ov 2}\,\frac{\sinh\frac{N\beta}{2}}{\sinh\frac{\beta}{2}} = \frac{1}{2\sinh\frac{\beta}{2}} e^{N\beta}-    \frac{1}{2\sinh\frac{\beta}{2}}\ . 
\ee
On the M-theory side this expression   is expected to   be reproduced  by the  M2 brane semiclassical contributions 
 of  the two saddle points:  of AdS$_{3,\b}$    corresponding 
 to M2 ending  on $S^1$   of  the $S^5$ boundary of AdS$_{7,\b}$   part of \rf{2}  (having non-zero classical  action)
 and of  a   degenerate  M2  brane  wrapping  only $S^1_\b$ (with zero action). 
 As we will show   below, the   fluctuation determinants   near the first  saddle point  reproduce precisely the 
   prefactor $[2\sinh\frac{\beta}{2}]^{-1}$   in \rf{2.13},   which is again in  full analogy with a similar computation in the 
   \adsz\   case in \ci{Giombi:2023vzu}.

\section{Semiclassical expansion of M2  brane  path integral }
\la{sec:setup}

Our aim   will be to consider a  semiclassical expansion  of the  Euclidean M2  brane  path integral near  
particular classical  solutions     in the ``twisted'' version  \rf{2},\rf{3} of the \adssf\ background. 
While the M2 brane action   \cite{Bergshoeff:1987cm}
  is  highly non-linear,  when  expanded near a classical  solution with a non-degenerate induced 
3d metric  it can be  straightforwardly quantized in a static gauge. 
Then  
 the  leading 1-loop  result for its partition function   is well defined (has no      UV   logarithmic   divergences)
\cite{Duff:1987cs,Bergshoeff:1987qx,Mezincescu:1987kj,Forste:1999yj,Drukker:2020swu,Giombi:2023vzu,Beccaria:2023ujc}.

The bosonic part of the M2   brane  action  may be written as 
\ba
&S = S_\V+S_{\rm WZ}, \la{31} \qquad \ \ 
S_\V = T_{2}\, \int d^{3}\xi\, \sqrt{g}, \ \ \ \qquad g_{ab} = \partial_{a}X^{M}\partial_{b}X^{N}\, G_{MN}(X), \\
&S_{\rm WZ} = -i\, T_{2}\, \int d^{3}\xi\, \frac{1}{3!}\eps^{abc}C_{MNK}(X)\, \partial_{a}X^{M}\partial_{b}X^{N}
\partial_{c}X^{K},  \qquad \ \ \  T_{2} = \frac{1}{(2\pi)^{2}\ell_{P}^{3}} \la{32}\ . 
%\\
%\la{3.9}
%\qquad \rt_{2} = a^{3}T_{2} = \frac{2}{\pi}N.
\ea
Here $S_\V$  is the  induced  volume (or Dirac-Nambu-Goto) term,  while $S_{\rm WZ}$ 
represents   the coupling to the $C_3$   potential  of 11d   supergravity. 
%%%%%%%%%
%%%%%%%%%%%%%%%%%%%%%
The  explicit   form of the fermionic part of the M2  brane  action  is also  known, in particular, 
 for the cases of  the maximally supersymmetric AdS$_4 \times S^7$
 or   AdS$_7 \times S^4$  backgrounds  
 \ci{Bergshoeff:1987dh,Bergshoeff:1988uc,Duff:1989ez,deWit:1998yu,
 Claus:1998fh,Pasti:1998tc}. It can 
  also  be found for the AdS$_{7,\b}  \times \tS^4$    background  \rf{2},\rf{3}  related to AdS$_7 \times S^4$  
  %AT16
   by an   ``orbifolding''   and  coordinate  redefinition. 
 The 1-loop  computation discussed below  will   require only the knowledge of  the quadratic fermionic term 
  in the M2 brane  action expanded  near a bosonic   background  $X^M(\xi)$
  \ci{Bergshoeff:1987dh,Bergshoeff:1987qx,Tseytlin:1996hs,deWit:1998tk,Russo:1998xv,Harvey:1999as}
    \ba
 & S_F =i T_2  \int d^3\xi\Big[ \sqrt g \,   g^{ab}   \del_a X^M  \,   \bar \theta \,  \Gamma_M \hat D_b \theta 
 - \ha \eps^{abc}  \del_a X^M  \del_b X^N  \,  \bar \theta \,  \Gamma_{MN}  \hat D_c \theta 
  + ...\Big]
  \ , \la{33} \\
  & g_{ab} = \del_a X^M  \del_b X^N   G_{MN} (X), \qquad \ \  \
   G_{MN} = E^A_M E^A_N, \qquad  \ \ \
  \Gamma_M  =   E_M^A(X) \Gamma_A\ ,    \la{334}
  \\
  &   \hat D_a = \del_a X^M   \hat D_M , \ \qquad 
  %v3 
  \hat D_M = {\rm D}_M  - \tfrac{1}{288} (\G^{PNKL}_{\ \ \ \ \ \   \ M} 
  +  8  \G^{PNK}\delta^L_M ) F_{PNKL} \ ,  \la{ff}
\ea
 where  $ \hat D_M $  is the   generalized 11d  spinor covariant derivative \ci{Cremmer:1979up} 
and  ${\rm D}_M=\del_M + \four \G_{AB} \omega^{AB}_M $.\foot{In the static gauge  $X^a= \xi^a, \ X^I=0$  ($I=1, ...,8$)
the natural $\k$-symmetry   gauge is like  in flat  space \ci{Bergshoeff:1987dh,Bergshoeff:1987qx}: \   $(1+ \G) \theta=0, 
\ \  \G= {1\ov 6 \sqrt g} \eps^{abc} \del_a X^M \del_b X^N \del_c X^K  \G_{MNK} $  or  alternatively 
 $(1 +  \G^{1...8}) \theta=0$.}
% is the standard  spinor covariant  derivative.

The action \rf{31}   computed  on the  twisted \adssf\ background  \rf{2},\rf{3}   depends on the   effective
dimensionless  M2  brane tension 
\be \la{3.9}
\rt_{2} = a^{3}T_{2} = \frac{2}{\pi}N \ . \ee
Thus  the  semiclassical    large   tension expansion of the M2  brane partition   function 
 should  correspond to the large $N$ expansion  on the  dual field theory side.

In general,  for an M2   solution   with a
 non-vanishing   classical    action  $S_{\rm cl} = \T_2\bar S_{\rm cl}$ 
 (where $\bar S_{\rm cl} $  represents  the total value of the sum of the volume  and  the WZ term  in \rf{31},\rf{32})
 the M2   brane partition function $\rm Z$  expanded near  this   background 
   will contain 
   a factor $ e^{-S_{\rm cl}}= e^{ -\T_2\bar S_{\rm cl}}=e^{- p N}$  where 
$p$   may depend on the  parameter $\b$ of the   background  \rf{2},\rf{3}.

Given an  M2 brane classical solution $X^{M}=X^{M}(\xi)$ \ ($M=1, \dots, 11$)
 we   may  choose  the  static gauge
 identifying three  of the $X^M$  coordinates with the M2 world-volume coordinates 
 $\xi^a \ (a=1,2,3)$  and also fix a  $\kappa$-symmetry gauge  for the fermions. 
   Then the  %  fluctuation  determinants   for the 
    remaining 8 bosonic and 8 fermionic   fluctuations
      will produce a $\b$-dependent   1-loop  prefactor  in the M2 brane   partition function $\rZ$ 
      %   The 1-loop  contribution  to the M2-brane partition  function  is then given by 
\ba \la{36}
&\rZ =\int [dX\, d\theta]\, e^{- S[X,\theta]} = \  \Ze_1\,  e^{-\T_2 \bar S_{\rm cl}}\big[1 + \OO (\T_2^{-1} )\big]   \ , \ \ \ \ \ \ \ \ \ \ \ 
S_{\rm cl} = \T_2 \, \bar S_{\rm cl}  \ , \
\\
%A19
&\Ze_1 = e^{-\G_1} \ , \ \ \ \qquad \   \G_1 =\ha \sum_k  \nu_k  \log \det \Delta_k
\ ,\la{zzz} \ea
where %$\hat V$   is the total value of the volume plus WZ term in \rf{31},\rf{32} 
$\Delta_k$ are  2nd-derivative fluctuation operators and $\nu_k=\pm 1$ for the  bosons and fermions.

    Below we will    consider  the  M2 branes  wrapped on  $S^1_\b$.
    The leading non-perturbative  $e^{-N\b}$ term in  the free energy \rf{2.10}   will be   reproduced 
     by the    solution that   wraps  also  the $S^2$ in the $\tS^4$   part of the metric in \rf{2}.
   %   It is 
       We will  also      consider the  solution   that  wraps an AdS$_{3,\b}$  part of AdS$_{7,\b}$  in \rf{2} 
       (ending on  the  big  circle of $S^5$)  
       that will reproduce the leading $e^{N \b}$   term in the Wilson loop expectation value in \rf{2.13}. %This   again is  in direct analogy to  the \adsz\    case in \ci{Giombi:2023vzu}). 

%%%%%%%%%%%%%%%%%%%%%%%%%%%%%%%%%%%%%%%%%%

\section{$S^{1}_{\beta}\times S^{2}$    M2  solution:   matching   non-perturbative free energy} %  term in $F$} 
% free energy}  %, with $S^{2}\subset S^{4}$}
\la{sec:calc-free}

Let us consider  the classical    M2  brane solution that is wrapped on $S^1_\beta$ in $\AdS_{7,\b}$   and $S^2$ in the $\tS^4$ part 
of the metric \rf{2}. 
 It is    an analog  of the  instanton M2 brane in \adsz\    discussed in \ci{Beccaria:2023ujc}. 
 Explicitly,  we may choose  the   coordinate $y$ in \rf{2}  to be $  \xi^3$  (assuming now that  $\xi^3 \in (0, \b)$)
 and  the coordinates of the unit-radius   $S^2 $  to be $\xi^1$ and $\xi^2$, with %  (with standard periodicities) 
 %M1 : I would remove ``(with standard periodicities)'', as these are just coordinates and also not necessarily angles, as in stereog.
  the rest of the coordinates in \rf{2}  being trivial, i.e.  $x=0, \ u=0$, etc.\foot{Keeping a  general  constant value of the 
 coordinate $u$  and computing the classical action   one can check that $u=0$  is an extremum.  Note also that the shift of 
 $z$ by $iy$ in \rf{2} is irrelevant at the classical level    at  the $u=0$ point.}

The  corresponding value
 of the classical  M2 brane  action in \rf{31} (cf. \rf{3.9})\foot{Here we introduced for convenience the notation $\rr$ for the relative factor $1\ov 2$ between the  radii  of AdS$_7$ and $S^4$ metrics in \rf{1}  and \rf{2}.}
  is then\foot{Note
  %v4
 that the same result  for the WZ term 
    is found  by  defining it as  $ - i T_2  \int  F_4 $      where integral goes over 
a space with topology of  half of 4-sphere (cf., e.g.,  \ci{Drukker:2005kx}).
With  this definition  the result  does not depend on  "large" gauge  transformations of  $C_3$
\ci{Kalkkinen:2003gq}  or  a   particular  gauge choice  made in   \rf{3}.
}
\ba
&S_{\V, \rm cl} = T_{2}\, a^{3}\rr^{2}  \vol({S^1_\b \times S^{2}}) = {1\ov 4} \T_2     \, \beta\, 4\pi = 2N\beta\ ,  \qquad \qquad \rr\equiv  \frac{1}{2} \ , \la{41}
\\
&S_{\rm WZ, cl} = -i\, T_{2}\, \int C_{3} =-\frac{1}{8}\, T_{2}a^{3}\, \int dy\wedge \vol_{S^{2}} = -\frac{1}{8}\, \rt_{2}\, \beta\, 4\pi = - N\beta\ , 
\la{42} \\
&S_{\rm cl} = S_{\V,\rm cl}+S_{\rm WZ,cl} = N\beta \ . \la{420} 
\ea
Thus $e^{-S_{\rm cl}}$   matches    the  exponential  factor
 in the leading term in the non-perturbative part  of free energy (\ref{2.10}).\foot{A similar   computation  in  the type IIA
 string limit (i.e. $\beta \to 0  $ with $\xi = \b N$=fixed) 
 was done in \ci{Gautason:2023igo}.  Wrapping M2 on  2-sphere 
     $n$  times we get a ``multi-instanton"    contribution  $S_{\rm cl} = n N\beta$  and thus   may match 
     the subleading $e^{- n N \b}$   terms
       in the  free energy  $F^{\rm np}$ in \rf{2.10}.
 Note that if we consider an ``anti-instanton'' 
   solution with 
  reversed  orientation of the $S^2 \to S^2$ map
 the contribution   of the $C_3$  term \rf{32}   in the action  will  then   have the opposite sign and thus we will get 
   $S_{\rm cl} = 2 N\beta + N\beta  = 3 N\beta $.
   %A19
   This  ``anti-instanton''   solution  should not be supersymmetric and thus presumably   should not be contributing 
   to the  free energy (we thank the authors of \ci{Gautason:2023igo}  for this  suggestion). 
   }
%  At the same time, the subleading  term  in free energy  $F^{\rm np}$ in \rf{2.10}  is of order $e^{-2\b N}$ 
 %  and thus should be  captured by some other  M2  brane  instanton  solution.} 
%If $u$ is left arbitrary, one can check that $u=0$ is an extremum.

\subsection{Quadratic fluctuation Lagrangian}

To discuss fluctuations near this classical solution we will choose  a   natural static gauge, i.e. set the fluctuations of  $y$ and $S^2$  coordinates to be zero. 
Let us first  discuss fluctuations in the $\AdS_{7,\b}$   directions  of \rf{2}  parametrizing  its metric as 
%\subsection{$AdS_{7}$ fluctuations}
\be
ds^{2}_{\AdS_{7,\b}} = \frac{(1+\frac{1}{4}{\chi}^{2})^{2}}{(1-\frac{1}{4}{\chi}^{2})^{2}}\, dy^{2}+\frac{d{\chi}^{p} d \chi^p}{(1-\frac{1}{4}{\chi}^{2})^{2}},
\qquad {\chi}^p = (\chi_{1}, \dots, \chi_{6}), \qquad y\equiv y+\beta.\la{44}
\ee
In  the static gauge $y=\xi_{3}$   the  6  fluctuations $\chi_p$ are thus functions of $\xi^3\equiv  \xi^3 + \b $ and the unit 2-sphere  coordinates.
As $C_3$  in \rf{3}   does not  involve AdS$_{7,\b}$ coordinates, we need to consider the quadratic fluctuation term of the 
$S_V$ part  of the M2  brane action \rf{31} only. 
Let   us  introduce the notation $g_{ij}$  ($i,j=1,2$)  for the unit-radius $S^2$   metric  so that the 3d  induced metric may be written as 
%A16
\be  ds^2= g_{ab} d \xi^a d \xi^b  =  g_{ij}(\xi)   d\xi^i d \xi^j +  d \xi^3  d \xi^3  \ , \ \ \ \ \ \ \ \ \ 
 g_{ij}(\xi)   d\xi^i d \xi^j = d \xi_1^2   + \sin^2 \xi_1\,  d \xi_2^2\ . \la{sss}
 \ee
Then expanding  to quadratic order in $\chi_p$ we get 
\ba
&S_\V = \rt_{2} \rr^2 \int d^{3}\xi\, \sqrt{ g}\, \big(1+\mathscr L_{2,\V}+\cdots\big),\la{45}
\\ 
&\mathscr L_{2,\V} (\chi) %&= \frac{1}{2}(\partial_{1}\chi^{p})^{2}+\sum_{a,b=2}^{3}\frac{1}{2}g^{ab}_{2,\rr}g^{ab}_{2,\rr}\, \partial_{a}\chi_{i}\partial_{b}\chi_{i}+\frac{1}{2}\chi_{i}\chi_{i}\lp
= \frac{1}{2\rr^{2}}\Big[ %\sum_{a,b=2}^{3}
 g^{ij}\, \partial_{i}\chi^{p}\partial_{j}\chi^{p}+{\rr^{2}}{}\chi^{p}\chi^{p}+ 
{\rr^{2}}(\partial_{3}\chi^{p})^{2}\Big]\  . \la{46}
\ea
The overall factor $1\ov \rr^2$ here  can be rescaled away   by redefining $\chi^p$.
Expanding $\chi_p$ in Fourier modes in the periodic  $\xi^3$  coordinate
we get   an  equivalent  2d    theory on  $S^2$  for a  tower  of  6 scalar fields  $\chi^p_n$   with masses  ($\del_3 \to i {2\pi \ov \b} n $) 
\be
\la{4.8}
M^{2}_{\chi,n}  = \rr^{2}(1+n_{_{\beta}}^{2})= {1\ov 4}   + {1\ov 4 }  n_{_{\beta}}^{2}\
,\qquad \qquad  n_{_{\beta}} \equiv  \frac{2\pi n}{\beta} \ , \qquad n= 0, \pm 1, \pm 2, ...\ . 
\ee
The remaining 2  fluctuations  in $\tS^4$ directions of \rf{2}  correspond to  $u$ and $z$
coordinates    which  represent  a 2-sphere  subspace $du^2 + \sin^2 u \, dz^2$.  
%(the shift of $z$ by $iy$ in \rf{2} is irrelevant at quadratic fluctuation level). 
Using   the Cartesian parametrization for this 2-sphere\foot{Explicitly, 
$ u = \text{arccos}\, { [1  - \frac{1}{4} (A^2 + B^2)]^2\ov  [1  + \frac{1}{4} (A^2 + B^2)]^2} ,
 \ \  z = \arctan  {B \ov A} $.}
\ba
du^2 + \sin^2 u \, dz^2 = { dA^2  +  d B^2 \ov [1  + \frac{1}{4} (A^2 + B^2)]^2 }\ ,  \la{49}
\ea
we  may use  $A$ and $B$  as  the two fluctuation  fields.   Rescaling  them by $(\T_2)^{1/2}$  (cf. \rf{45}) 
we then get  the following counterpart of \rf{46}  coming from the  volume part of the M2 brane action in \rf{31} 
%%%%%%%%%%%%%%
\ba
\la{4.12}
%S_{2}|_\V &= \int d^{3}\xi\, \sqrt{G_{S^{1}\times S^{2}_{\rr}}}\, \mathscr L_{2}|_\V, \\
\mathscr L_{2,\V} (A,B)
%&= \frac{\rr^{2}}{2}G^{ab}_{S_{1}\times S^{2}_{\rr}}(\partial_{a}A\partial_{b}A+\partial_{a}B\partial_{b}B) 
%-\frac{\rr^{2}+2}{2}(A^{2}+B^{2})+2i\, \rr^{2}A\partial_{1}B \lp
=&\frac{1}{2}\Big[
g^{ij}(\partial_{i}A\partial_{j}A+\partial_{i}B\partial_{j}B) \no \\
&\ \ \  +  {\rr^{2}}(\partial_{3}A)^{2}+{\rr^{2}}(\partial_{3}B)^{2}
-  (\rr^{2}+2)(A^{2}+B^{2})+4i\,\rr^{2}A\partial_{3}B \Big] \ .
\ea
Here the mixing term $A\partial_{3}B$   is   due to the presence of $dy=d\xi^{3}$ in the  $(dz+ idy)^{2}$ term in \rf{2}
(cf. \ci{Russo:1998xv}).
%Notice that we have factored
%\be \sqrt{G_{S^{1}\times S^{2}_{\rr}}} = \rr^{2}\, \sqrt{G_{S^{1}\times S^{2}}}. \ee

%\medskip\noindent
For the contribution of the WZ  term  in \rf{32} with $C_3$ in \rf{3} one finds using that $\del_3 y=1$   (cf. \rf{42})   %Evaluating the WZ term, one obtains 
\ba
S_{\rm WZ} &= -iT_{2} \int C_{3} = \frac{i}{8}\rt_{2}\int \cos^{3}u\,(\partial_{3}z + i )\,d\xi_{3}\wedge \vol_{S^{2}} 
%= -\frac{1}{8}\rt_{2}\int d^{3}\xi\,\sqrt{G_{S^{1}\times S^{2}}}\ \cos^{3}u\, (1-i\partial_{1}z) 
= -\frac{1}{8}\rt_{2}\int d^{3}\xi\,\sqrt{g}\ \cos^{3}u\, (1-i \partial_{3}z ).  \la{411}
\ea
Expanding to  quadratic order in the fluctuations $A,B$  we get the following addition to \rf{4.12}
\ba
\la{4.16}
\mathscr L_{2,\rm WZ} (A,B)  = \frac{3}{16\rr^{2}}(A^{2}+B^{2})-\frac{3}{8\rr^{2}}i\, A\partial_{3}B\  .
\ea
%%%%%%%%%%%%%%%%%%%%%%%%%
%%%%%%%%%%%%%%%%%%
Summing up \rf{4.12}  and \rf{4.16}   gives (setting $\rr=\ha$    and ignoring a  total derivative)
\ba
&\mathscr L_{2} (A,B) = \frac{1}{2}
g^{ij}(\partial_{i}A\partial_{j}A+\partial_{i}B\partial_{j}B)   +  \mathscr L_{2, M}(A,B) \ , \qquad 
\la{413} \\ &
\mathscr L_{2, M}(A,B)
= -\frac{3}{8}(A^{2}+B^{2}) +\frac{1}{8}\big[(\partial_{3}A)^{2}+(\partial_{3}B)^{2}\big]-i\, A\partial_{3}B
\ .\la{414}  \ea 
Setting  $ \phi = \frac{A+i B}{\sqrt 2}, \  \bar\phi = \frac{A-i B}{\sqrt 2}$   we get 
\ba
\mathscr L_{2, M}(\phi)
= \frac{1}{2}\begin{pmatrix} \phi &  \bar\phi \end{pmatrix}
\begin{pmatrix} 0 & -\frac{3}{4}+\partial_{3}-\frac{1}{4}\partial_{3}^{2} \\
 -\frac{3}{4}-\partial_{3}-\frac{1}{4}\partial_{3}^{2} & 0  \end{pmatrix}
\begin{pmatrix} \phi \\  \bar\phi \end{pmatrix}\ . 
 \la{4133}\ea
Expanding $\phi (\xi)$ in Fourier modes   in $\xi^3$   we get an effective 2d  Lagrangian  for a 
  tower of complex scalars  on $S^2$ (cf. \rf{4.8}) 
\ba
&\mathscr L_{2}(\phi)  =\sum^\infty_{n=-\infty} \big(  g^{ij}\partial_{i}\phi_n \partial_{j}\bar \phi_{n}  + M^2_{\phi,n}  \phi_n \bar \phi_{n} \big) \ ,\la{641}\\ 
&M^{2}_{\phi, n} = -\frac{3}{4}+i n_{_{\beta}}+{1\ov 4} {n_{_{\beta}}^{2}}=  1 +  {1\ov 4} (\nb + 2 i)^2 \ .    \la{416}
\ea
 In the limit  $\beta\to 0$  the current problem  should  reduce to  the type IIA   string computation  considered  in 
 \cite{Gautason:2023igo}:   the string spectrum   should   be  the $n=0$   level  of the   M2  brane spectrum. 
 Indeed, the $n=0$  values  of the masses  of the  6   fluctuations   in \rf{4.8} and 
 2    fluctuations  in \rf{416} agree  with the  bosonic  string fluctuation masses in 
 Table 1 of \cite{Gautason:2023igo}.

 %A19
The   fermionic   part  of the M2  brane action directly   corresponds  (upon double dimensional reduction  as in 
\ci{Duff:1987bx})  to the  fermionic part of  type IIA superstring  action.  In  the superstring   limit  one finds 
 \cite{Gautason:2023igo}   that the quadratic part of the GS  action   is equivalent  to 
    8 fermions  in  $S^2$  geometry  with the square of the Dirac operator containing the  mass term with 
      $M^2= - {1\ov 4}$.  Explicitly, the   2d Dirac  operator  is given by (cf.  \cite{Gautason:2023igo,Beccaria:2023ujc}):  
${\cal D} = i \sigma^k {\rm D}_k + M \s_3 $  where 
 $\sigma_a$ are the  three   Pauli  matrices   with  the $\s_3$ term originating from the  terms   with $\G_{11}$  factors 
  in the membrane   action \rf{33}.  
  Its square is $\Delta_\ha  = - {\rm D}^2 + {1\ov 4} R^{(2)}   + M^2$, where $R^{(2)} = 2$  is the curvature of the 
   2-sphere. In the type IIA string limit     \cite{Gautason:2023igo}  one gets     $M=-\ha  i $. 
   
 Starting  directly with the M2  brane action \rf{33},  %     and   accounting for the dependence on the  $\xi^3=y= x^{11} $ 
in the present case  with $y= x^{11}= \xi^3$  
  there  are  two  different  $M \s_3$ contributions to the fermionic ${\cal D}$ operator. 
  One  is     coming from the non-zero $y$-component of $F_4$ 
  field   strength corresponding to $C_3$  in \rf{3}   that gets   contracted with $\G^y$ or $\G^{11}$ in  \rf{ff}. 
 This    corresponds   upon double dimensional reduction to  a   similar term  in the  type IIA string action
 leading precisely to  the above  $-\ha  i $ contribution to $M$.  
% comparing to \cite{Gautason:2023igo} this  corresponds to the above   $\ha i$ contribution to $M$.    
 %     In the  membrane   action,  expanding in Fourier modes in $\xi^3$ we 
%      will   should  a  tower of 2d fermionic modes on $S^2$  with $M^2$  that 
  %   get extra $ c_1  \nb + c_2  \nb^2$ contributions. 
     
 In addition, there is  a contribution of the $\Gamma_{11}  \partial_{3}$  term  in the covariant derivative in \rf{33},\rf{ff}
  (cf.  also Eq.~(4.29) in \cite{Beccaria:2023ujc}).\foot{To find the quadratic fermionic  term in the 
    M2  brane  action what matters is  the form of the 
     classical bosonic  $X^M(\xi)$  background  that 
gives the induced  3-bein  contracted with $\Gamma_M$. In
 the present case this  is coming from the $y$-dependent  terms in the metric \rf{2}.  
%$ds^2 = \dots + \cosh^2 x dy^2 +   \frac{1}{4}(\dots  - \sin^2 u dy^2   + \dots)$.
On the classical solution $u=0$, $x =0$, $y= \xi^{1}$,
 the only term that    is relevant  originates  simply from the    $dy^2$  term in \rf{2}. }  
 This  gives  an extra $-\ha \nb$  contribution  to the fermion mass  $M$  so that in total 
 $M= - \ha \nb - \ha i $. 
 
 As a result,   
 we get 8 towers of 2d  fermions on $S^2$   with % masses
 \be \la{417}
 M^2_{\theta, n} = \frac{1}{4} ( \nb + i)^2 = -  \frac{1}{4}  + {i\ov 2} \nb  + \frac{1}{4} \nb^2 \ . \ee
Combined  with the 6+2   towers of bosons in \rf{4.8}  and \rf{416} this represents  the complete 
M2 brane   fluctuation spectrum.

\subsection{One-loop  M2   brane partition function}

The expressions  for  the  determinants of  the standard  bosonic  and fermionic  massive field operators  on $S^{2}$
in the 1-loop  contribution in \rf{zzz}  are well known. 
In general, using  spectral zeta-function regularization  one has 
 $\log \det \Delta = - \zeta_\Delta(0) \log \Lambda^2  - \zeta'_\Delta(0)$. 
Like in \cite{Giombi:2023vzu,Beccaria:2023ujc}    the total 
  coefficient $\zeta_\Delta(0)$ 
  of the log UV divergence  vanishes  if we use  the  Riemann zeta-function regularization of the sum over the 
  modes  (that 
  removes power divergences) %(reflecting the   fact  that  we are dealing with a 3d theory):
  \be
\la{4.27}
\zeta_{\rm tot}(0) = \sum_{n\in \Z} 2 = 2  + 4 \zeta_R (0) = 0\ .
\ee
Here the coefficient 2 is related to the   value of the  Euler number of $S^2$ (cf. \ci{Giombi:2020mhz}). 
The finite $ -\zeta'_\Delta(0) $ parts of $\log\det \Delta $  for the bosonic  ($\Delta_0 = - D^2 + M^2$) 
and  fermionic  ($\Delta_\ha  = - {\rm D}^2 + {1\ov 2} +  M^2$)  fields  on $S^2$  are given by 
(we  follow the notation in  \cite{Gautason:2023igo,Beccaria:2023ujc})
 \ba\la{428}
& \log\det \Delta_0  = s_{\frac{1}{2}}(\tfrac{1}{4}-M^{2}) \ , \qquad \ \ \ 
  \log\det \Delta_{1\ov 2}   = s_{0}(-M^{2}) \ , \\
 &\la{4.26} 
 s_{p}(\mu) \equiv  -4\zeta'(-1,p)+\int_{0}^{\mu}dx\ \Big[\psi(p+\sqrt x)+\psi(p-\sqrt x)\Big]\ , 
\ea
where  
$\zeta'(x,a)$  is the   derivative of the   Hurwitz   $\zeta$-function  over $x$
and    $\psi$  is the logarithmic derivative of the $\Gamma$-function. 

As a result,  combining together the  contributions  of the above (6+2)  bosonic and 8   fermionic determinants 
and summing over $n$ we  find 
\ba\la{422}
 \Gamma_{1} = \sum_{n\in\mathbb Z}U\Big(\frac{2\pi n}{\beta}\Big),\qquad \ \ 
U(\nu) =  3 s_{\frac{1}{2}}\big(-\tfrac{\nu^{2}}{4}\big)+s_{\frac{1}{2}}\big((1-\tfrac{i\nu}{2})^{2}\big)
-4 s_{0}\big((\tfrac{1}{2}-\tfrac{i\nu}{2})^{2}\big)\ .
\ea 
% 3 s_{\frac{1}{2}}\Big(-\tfrac{\nu^{2}}{4}\Big)+s_{\frac{1}{2}}\Big(1-c_{1}\nu-\tfrac{\nu^{2}}{4}\Big)
%-4 s_{0}\Big(\tfrac{1}{4}-c_{2}\nu-\tfrac{\nu^{2}}{4}\Big). \ea Using $c_{1}=i$ and $c_{2}=\frac{1}{2}c_{1}$, this reads 
Using  (\ref{4.26})  we  observe that\footnote{Recall  that $\zeta_R'(-1)=\frac{1}{12}-\log\mathsf{A}$ and $\zeta'(-1,\ha) = -\frac{1}{24}-\frac{1}{24}\log 2+\frac{1}{2}\log\mathsf{A}$  where 
 $\mathsf{A}$ is Glaisher's constant.}
\ba U(0) = i\pi-2\log 2 \ ,  \qquad \qquad 
U(\nu)+U(-\nu)  &= -4\log 2 +2 \log(1+\nu^{2}), \qquad \nu>0 \ .\la{423}
\ea
Thus, like in the  case of the instanton M2 brane  solution in \adsz\ in   \cite{Beccaria:2023ujc}, 
   all non-trivial  $\psi$-function dependent  terms  from  \rf{4.26} 
cancel out   in the sum of the bosonic and fermionic contributions\foot{Note that these cancellations 
 would not happen if we were to  ignore the $z\to z + i y$ twist in the metric \rf{2}  which appears to   be  consistence 
 %A16
  with its   need  for preservation of supersymmetry.} 
and  we  end up with % the following expression
\ba
\Gamma_{1} &= i\pi-2\log 2+\sum_{n=1}^{\infty}\Big[-4\log 2+2 \log\Big(1+\frac{4\pi^{2}n^{2}}{\beta^{2}}\Big)\Big] \lp
= i\pi-2\log 2\big(1 + 2\,\zeta_R(0)\big) +2\log\big(2\sinh\frac{\beta}{2}\big) = \log\big(-4\sinh^{2}\frac{\beta}{2}\big).
\la{424}
\ea
As a result,  the 1-loop  factor in the  M2-brane  partition function \rf{36} 
on this  M2  instanton background 
 is given by 
\be\la{425} 
\Ze_1 = e^{-\Gamma_{1}} = -\frac{1}{4\sinh^{2}\frac{\beta}{2}} \ . 
\ee
Taking into account  that,   as  discussed in \cite{Beccaria:2023ujc},    
 the field-theory  free   energy   should   be  matched by   {\it minus} 
the M2  brane     partition function,  we  thus reproduce 
 the prefactor in the leading 
non-perturbative term in the  free energy  in (\ref{2.10}).
This generalizes   to finite $\b$ case the matching found   in   the string theory  limit  in 
\ci{Gautason:2023igo}.

\section{$\AdS_{3,{\beta}}$  M2  solution:   matching  Wilson loop expectation value } % $\langle W \rangle$} 
% free energy}  %, with $S^{2}\subset S^{4}$}
%\la{sec:calc-free}
%\section{Wilson loop: M2 brane in $\AdS_{2}\times S^{1}_{\beta}\subset AdS_{7}$}
\la{sec:calc-wl}

In analogy with the   discussion in the \adsz\  case in  \ci{Giombi:2023vzu},   the leading term in the  circular BPS 
  Wilson loop  expectation  value in \rf{2.13}  is  to be reproduced  by  the  M2  brane  
  partition function  expanded near  the 
   solution that ends on a circle 
  of the $S^5$ part of the  boundary  and is also 
  wrapped on the 11d  $y$-circle of $\AdS_{7,\b}$ while 
  being point-like in  $\tS^4$ part of  in  \rf{2}.
   This    should   generalize to the  M-theory  (finite $\b$)   level  the   related computation done in the
    type IIA string theory limit in \ci{Gautason:2021vfc}. 
    
 Denoting   by $\vp\equiv \vp+2\pi $   the circular coordinate of $S^5$,   the  relevant  $\AdS_{3,\beta}\subset \AdS_{7,\beta}$  part of the 
 metric \rf{2}   and thus the induced metric  for   the classical M2   solution  $x=\xi^1, \ \vp= \xi^2, \ y= \xi^3\equiv \xi^3 + \b$
 will be  that  of  ``thermal'' $\AdS_{3,\beta}$
 \be ds^2_{\AdS_{3,\beta}}= dx^2 + \sinh^2 x\,  d\vp^2 + \cosh^2x\,  dy^2  \ \to \   g_{ab}(\xi) d \xi^a  d \xi^b = 
 d\xi_1^2 + \sinh^2 \xi_1\,  d\xi_2^2 + \cosh^2\xi_1\,  d\xi_3^2\ .   \la{51} \ee
 %with the classical M2   solution being $x=\xi^1, \ \vp= \xi^2, \ y= \xi^3$. 
  The corresponding  classical M2  brane action gets only the volume  contribution \rf{31}, i.e. 
  \be\la{52}
    S_{\V,\rm cl} = \T_2 \vol({{\AdS_{3,\beta}}}) = - N \b \ . 
  \ee
The  computation of the regularized   volume  of ${\AdS_{2n+1,\beta}}$ with boundary $S^{2n-1} \times S^1_\b$ is  reviewed in Appendix 
\ref{app:vol}. Explicitly, 
\be  \vol({{\AdS_{3,\beta}}}) = \int_{0}^{\beta}dy\ \int_{0}^{2\pi}d\vp\ \int_{0}^{x_{0}}dx\, \sinh x\cosh x = 
\beta\,\pi\,\sinh^{2}x_{0}  = {1\ov 4}  \pi \beta \big( {1\ov  \eps^2 } -2  + \eps^2\big) \to -\ha \pi \beta \ ,   \la{520} 
\ee
where  we set  $x_{0}=-\log\eps $  as  IR cutoff  $(\eps\to 0) $   and   dropped power divergence. Using  \rf{3.9}  we thus get
the value in  \rf{52}  which indeed matches  the exponent of the first term in \rf{2.13}
(see also   \cite{Minahan:2013jwa}).  
 The   second term in \rf{2.13}  may be expected to come from an M2 brane  solution with
    vanishing 3-volume  but this remains to be clarified.

\subsection{Quadratic fluctuation Lagrangian}

Choosing the static   gauge in which the fluctuations of $x,\ \vp$ and $y$ are   set to zero 
one can check (see below)   that   since the classical   solution is  trivial  in the $\tS^4$ directions, 
 the only   contribution to the 
quadratic fluctuation  action  comes   from the volume part \rf{31} of the M2  brane action. 

The   part of the quadratic  fluctuation Lagrangian depending only on the AdS$_{7,\b}$ coordinates in  \rf{2}
 is represented by the four     $S^5$ directions   that have trivial classical values.
%   ($C_3$ term in \rf{32}   does not contribute  to quadratic  fluctuation order as the $\tS^4$ 
Parametrizing the  $S^{5}$  metric as\foot{Same result for quadratic fluctuations 
 is found  if we use the Hopf fibration parametrization of the 
$S^5$ metric, i.e. $dS_{5}=  (d \vp'  + A)^2 +   ds^2_{\CP^2}$  where $A$  depends on  $\CP^2$  coordinates.}
 
\be\la{54}
dS_{5} = \frac{(1-\frac{1}{4}{w}^{2})^{2}}{(1+\frac{1}{4}{w}^{2})^{2}}\, d\vp^{2}+\frac{dw_{r}dw_{r}}{(1+\frac{1}{4}{w}^{2})^{2}}, \qquad
\qquad  r = 1, \dots, 4 \ , 
\ee
and expanding in powers of $w_r$   we get  from \rf{31} (we rescale away the overall factor of tension) 
 \be\la{55}
S_{2,\V}(w)  =\frac{1}{2} \int  d^3\xi  \sqrt{g}   g^{ab}   \sinh^2 \xi_{1} \big(   \partial_a w_r  \partial_b  w_r   -     \delta_{a 2} \delta_{b2}  w^{2}
\big)\ .
\ee
Setting 
\be
w_r =  \frac{1}{\sinh\xi_{1}}\wt{w}_{r} \  ,
\ee
and integrating by parts   
 we get 
\ba
\la{5.14}
S_{2,\V}(\wt w) &=  \int d^{3}\xi\, \sqrt{g}\, \mathscr L_{2,\V} (\wt w) \ , \qquad \qquad  
\mathscr L_{2,\V}(\wt w)
 = \frac{1}{2} \big( g^{ab} \partial_{a}\wt w_{r}\ \partial_{b}\wt w_{r}+3 \wt w_{r}\wt w_{r} \big)\ .
\ea
To find   the  contribution of the other 4  bosonic fluctuations   corresponding to   $\tS^4$ 
directions  in \rf{2}  we note that  the  leading part of the $\tS^4$ 
 metric  expanded   near $u=0$ 
  is    $\four [ du^{2}+ dS_{2}+u^2(dz + i \, d\xi_{3})^{2}] $.
  Using  Cartesian   coordinates $(A,B)$   to parametrize  the $(u,z)$ plane and $v_k \ (k=1,2)$ for  $S^2$, i.e. 
\be
\la{5.17}
A= u\cos z\ , \ \ \  B= u\sin z \ ,\ \ \ \qquad dS_{2} =\frac{dv_k dv_k}{(1+\frac{1}{4} v^{2})^{2}}\ , 
\ee
we   get the quadratic  fluctuation Lagrangian (rescaling all 4 fields by factor of $\rr={1\ov2}$; here $i,j=1,2$)\foot{Explicitly, we use that 
 $\int d^{3}\xi\, \sinh\xi_{1}\cosh\xi_{1} \frac{1}{\cosh^{2}\xi_{1}}[(\partial_{3} A)^{2}+(\partial_{3}B)^{2}-A^{2}-B^{2}
+2i (A \partial_{3}B-B\partial_{3}A)]
= \int d^{3}\xi\, \sqrt{g}\, g^{33}\big[(\partial_{3} A-i B)^{2}+(\partial_{3}B+iA)^{2}\big]$.}
\ba
&{\mathscr L}_{2,\V} (v_k,A,B)
 =\frac{1}{2}g^{ab
 }\big(\del_a v_k \del_b v_k + \del_a A \del_b A  +\del_a B \del_b B  \big)
       -\frac{1}{2}\frac{1}{\cosh^{2}\xi_{1}}   (A^{2}+B^{2})+\frac{2i}{\cosh^{2}\xi_{1}}\ A \partial_{3}B\no \\
&    = \frac{1}{2}\Big( g^{ab} \del_a v_k \del_b v_k +  g^{ij} ( \del_i A \del_j A  +\del_i B \del_j B) 
  +    g^{33} [(\partial_{3} A-i B)^{2}+(\partial_{3}B+iA)^{2}]\Big),\la{518}
\ea
Note that the $(A,B)$ mixing term  may  be formally  diagonalized by a  $\xi^3$-dependent  ``rotation''  %with angle $\psi=i\xi_{3}$. Indeed, setting
\ba
&A = \cos \psi\,  X  +   \sin \psi\,   Y, \qquad 
B = - \sin \psi\, X + \cos  \psi\,  Y , \qquad \psi=i\xi^{3} \ , \la{513} \\
&  (\partial_{3} A-i B)^{2}+(\partial_{3}B+iA)^{2} = (\partial_{3}X)^{2}+(\partial_{3}Y)^{2}. \la{511}
\ea
Since $\xi^3$  is periodic, this redefinition  is only formal 
as it    shifts  the  value of  the  $S^1_\b$  mode number  (cf. \rf{4.8})
as $\nb\to \nb  +  i$   and 
this  should be taken into account.

%A16
Indeed,  here we have a  coupling  of the  complex  scalar  $A+i B$ 
to a 
 constant 3d gauge  potential  with the  component 
  $\A_3 = - i$
  in  the $S^1_\b$  direction   which   can not be gauged away.\foot{Equivalently,   this is the $SO(2)$ 
   gauge field  coupled to $\Phi_k=(A,B)$ via $ D_3 \Phi_k = \del_3 \Phi_k + \eps_{kl} \A_3 \Phi_l $, cf. \ci{Russo:1998xv}.} 
   Its  origin is related to the  presence of the   twist  $z\to z + i y$  in the  metric \rf{2}. 
%Indeed,  if  we expanding in Fourier modes  in $\xi^3$  we get  an effective 2d theory 
%%%%%%%
%%%
This shift is similar to what we found in  the $\tS^{4}$ part  of the fluctuation Lagrangian \rf{416} in the $S^2$ instanton case 
where $n_{_{\beta}} + 2i$ rather than $n_{_{\beta}} + i$ was due to the contribution of the WZ term.\foot{Again, the origin of this  shift  can be   traced to the structure of the metric  in \rf{2}:  in view of the definition of $A,B$ in \rf{5.17},  redefining $z\to z + i  \xi_3$   translates into  the   rotation   (\ref{513}).}

Finally,   let us note that the $C_3$ coupling term \rf{32}  evaluated  on the background  \rf{3} 
gives
\ba
S_{\rm WZ} &= -iT_{2} \int C_{3} = \frac{i}{8}\rt_{2}\int \cos^{3}u\,(dz+ i dy)\,\wedge 
\frac{dv_{1}\wedge dv_{2}}{(1+\frac{1}{4}(v_{1}^{2}+v_{2}^{2}))^{2}}\ . \la{512}
\ea
Since $y=\xi^3$,   expanding 
 to quadratic order in the fields projected on the world-volume  
 this   reduces to a total derivative term $ \epsilon^{ij} \epsilon^{kl}\del_i v_{k} \del_j v_{l} $ 
  and thus   does not indeed contribute to the  leading order.

 As for the   fermionic  fluctuation  Lagrangian, it   can be found   by a  generalization 
 of its string theory ($\b\to 0$)   limit discussed in \ci{Gautason:2021vfc}. 
 We should   get 8    fermions in $\AdS_{3,\b}$ 
  with ${\cal D} = i \sigma^k {\rm D}_k + M \s_3 $  where  $M={3\ov 2}$. 
The    $\del_3$ derivative term  in ${\rm D}_k$  produces (upon Fourier expansion in $\x^3$)  a mode number $\nb$ contribution as in 
  \rf{4.8},\rf{417}.  % {\bf is there a flat connection here ?}
  Also,  as in the  case of the $(A,B)$ fields in \rf{518}, 
   here the covariant derivative  contains (in addition to   the standard $\AdS_{3,\b}$   spin connection) 
      a constant $U(1)$ potential  term, reflecting  again the presence of the twist in the metric \rf{2},  i.e. we have 
      (cf. \ci{Russo:1998xv})
   \be \la{526}
   {\rm D}_3= \del_3 -i   \A_3 + ... , \qquad    \A_3 = - \tfrac{1}{2} i  \ . \ee 
  %%%%%%%%%%%
\subsection{One-loop M2  brane partition function}
%Functional determinants for   fluctuations}
%\subsubsection{Bosonic case}

The  fluctuation Lagrangian  represents a collection of  massive bosons  and fermions propagating 
in $\AdS_{3,\b}$, i.e. in ``thermal'' AdS$_3$   with $S^1\times S^1_\b$ boundary. 
The expressions  for the corresponding determinants are  well-known from the literature (see, e.g, 
%A19
\cite{Gibbons:2006ij, Giombi:2008vd, Gopakumar:2011qs}). 

For a scalar field with mass $M$ %, the corresponding functional determinant is given  by Eq.~(4.10) in
one finds  \cite{Giombi:2008vd}\foot{This expression  was found in \cite{Giombi:2008vd} 
 (for the Casimir contribution see \cite{Giombi:2014yra}) 
 by applying the method of images to the heat kernel 
for the thermal quotient of $\AdS_{3}$. 
It is rederived in an alternative way  in Appendix \ref{app:s1s1}  below by  using the 
explicit  expansion  in modes 
along the two boundary circles $S^{1}\times S^{1}_{\beta}$, \cf (\ref{E.20}).}
\ba
\la{5.27}
&\G^{(\Delta)}(\beta) \equiv   \frac{1}{2}\log\det(-D^{2}+M^{2}) % \equiv \mathscr   D^{(\Delta)}(\beta) \ , \qquad 
= E_{c}(\Delta)\,\beta-  \sum_{n=1}^{\infty}\frac{e^{-\b n\Delta}}{n(1-e^{-\b n})^{2}}\ , %\equiv -  \mathscr  D^{(\Delta)}(\beta), 
\\
& % q=e^{-\beta}, \qquad\qquad  
\Delta = 1+\sqrt{1+M^2} \ . \la{514}
\ea
Here $\Delta$ is the conformal dimension of the ``dual boundary field'' and $E_{c}$ is the Casimir energy
\be\la{515}
E_c(\Delta) = \frac{1}{2\Gamma(z)} \int_0^{\infty} d\beta\ \beta^{z-1}\ \frac{e^{-\b \Delta}}{(1-e^{-\b})^{2}}\, \Big|_{z\to -1}
= \frac{1}{24}(\Delta-1)\,(1-4\Delta+2\Delta^{2})\ . 
% Z(\beta)\Big|_{z\to -1}, \qquad Z(\beta) = \frac{q^{\Delta}}{(1-q)^{2}}.
% and this evaluates (for a scalar field) to  $E_{c} = \zeta(-1,\Delta/2)$.
\ee
For $\beta\to \infty$ we have $\G^{(\Delta)}(\beta) = E_{c}(\Delta)\, \beta+\mc O(e^{-\b\Delta})$, 
while for $\beta\to 0$ one finds  (see Appendix \ref{app:smallbeta})
\ba
%\mathscr{D}^{(\Delta)}(\beta) 
\G^{(\Delta)}(\beta) = &- \frac{\zeta (3)}{\beta ^2}+ \frac{\pi ^2 (\Delta-1 )}{6 \beta }- \mathscr C(\Delta) +\frac{1}{12} (5-12 \Delta +6 \Delta ^2) \log \beta \lp
 +\frac{(1-20 \Delta +50 
\Delta ^2-40 \Delta ^3+10 \Delta ^4) }{2880} \beta ^2 + \OO(\b^4) \ ,  \la{5.31}\\
%+\frac{(-5+42  \Delta +63 \Delta ^2-420 \Delta ^3+525 \Delta ^4-252 \Delta ^5+42 
%\Delta ^6) \beta ^4}{3628800}+\cdots,
\mathscr C(\Delta) =  &(\Delta-1 )\Big [\frac{1}{2} \log (2 \pi )- \log\Gamma(\Delta )\Big]+\zeta'(-1,\Delta )\ .\la{5311}
\ea
We  still   need  to address  the following  subtlety: the   scalars $A,B$  in \rf{518}   are  not just  massless  scalars 
 in $\AdS_{3,\b}$ but are coupled also to a flat but topologically non-trivial $U(1)$ gauge potential in $\xi^3$ direction that 
 leads to a shift $\nb'= \nb + i$ of the $S^1_\b$   mode number. 
To account for the  effect   of such  coupling  on the scalar  determinant we may   use the   path integral representation for the 
log det  or heat kernel  of the  fluctuation operator  in \rf{518}   defined on the  complex  scalar $A + i B$ 
    in which the coupling to a background  3d 
gauge  field  $\A_a$ appears as a 
 phase factor $\exp [ i \int d\tau\,  \A\cdot\dot{x}]$. 
 For  constant 
 \be 
 \A_3=- i \k  \la{con}\ee
   this gives  a factor of $e^{ m   \kappa \beta}$ where $m$ is the number of times the worldline $x(\tau)$ wraps around the thermal circle (in \rf{518}  we have  $\k=1$).
%, then this should just add to the heat kernel a factor of the exponential 
%$e^{  \kappa \beta m}$ where $m$ is the number of times the worldline $x(\tau)$ wraps around the thermal circle. 
%In this representation the heat kernel is written as a sum over this wrapping number. 
This implies the following modification of (\ref{5.27})
\ba
\la{5.34}
\G^{(\Delta,\kappa)}(\beta) &=\ha \Big[  \G^{(\Delta+\k)}(\beta)  +  \G^{(\Delta-\k)}(\beta)\Big]  = 
 E_{c}(\Delta, \kappa) \beta
 - \frac{1}{2} \sum_{n=1}^{\infty}\frac{1}{n}\frac{e^{-\b n\Delta} ( e^{-\b\k} + e^{\b \k}) }{(1-e^{-\b n})^{2}}\ , \\
 E_{c}(\Delta, \kappa) &= \frac{1}{2}[E_{c}(\Delta+\kappa)+E_{c}(\Delta-\kappa)] = \frac{1}{24}(\Delta-1)(1-4\Delta+2\Delta^{2}+6\kappa^{2})\ . \la{519} 
\ea
%Note that this expression  even in $\k$. 
This   is 
derived directly  using  the $S^{1}\times S^{1}_{\beta}$  mode expansion 
 in Appendix \ref{app:s1s1}, see   (\ref{E.28}).

%\paragraph{Remark:} in the high temperature limit $\beta\to 0$, the singular terms $\sim 1/\beta^{2}$ and $1/\beta$ in (\ref{5.31}) are independent on $\kappa$.
% \subsubsection{Fermionic case}

The determinant   of the squared  massive    Dirac operator in  $\AdS_{3,\b}$, i.e.  
$\Delta_\ha  = - {\rm D}^2 + {1\ov 4} R^{(3)}   + M^2$, where $R^{(3)} = -6$,   
is given  by the same  expression  as in \rf{5.27}   %and  Appendix \ref{app:spinor})
but instead of  the relation between $\Delta$  and $M$  in  the scalar case in  \rf{514}  here   one has 
(see, e.g.,  
\cite{David:2009xg,Datta:2011za,Datta:2012gc,Kakkar:2023gzu})
% with  the  modified  relation between $\Delta$ and $M$ 
\be\la{5201} 
\Delta = 1+|M| \ . 
\ee
Eq. \rf{5201}   is the $d=3$ case of the standard AdS$_{d}$/CFT$_{d-1}$ relation 
for the fermions 
$\Delta = \frac{d-1}{2}+|M|$ (see, e.g., \cite{Henningson:1998cd}).\footnote{In general, 
for a spin $s$ field  in AdS$_3$ with the operator $-  {\rm D}^2_{s} + \mu^2$ one has  % (\eg \cite{Keeler:2019wsx},
$\Delta=1+\sqrt{\mu^2+s+1}$.
Thus for $s=\ha$  we get $\Delta = 1+\sqrt{\mu^{2}+\frac{3}{2}}$.
Since   here $\mu^2= {1\ov 4} R^{(3)}   + M^2= - {3\ov 2} + M^2$
we get  $\Delta = 1+|M|$.
}
The generalization to the case of the presence of  a   constant  gauge potential $\A_3=- i \k$
 is straightforward  as this  coupling is    via  the $\rm D_3$ term in the covariant derivative
  and thus the same as in the  scalar case. It is   given  again by  \rf{5.34}.

We are now ready to compute the   total contribution  to the 1-loop effective action \rf{zzz}
in the present case. According to the  discussion  in the previous  subsection  we have 
4 scalars with $M^2=3$  in \rf{5.14},  2 massless scalars $v_k$ in \rf{518}, 
 two   scalars $(A,B)$  in \rf{518},\rf{511}     with $M^2=0$  coupled to  a
  constant   potential \rf{con} with $\k=1$  and 8  fermions with $M=  {3\ov 2}$ 
  coupled to \rf{con} with $\k=\ha$ (see \rf{526}).\foot{Note that
  %v3 corr ???
    the corresponding values of $\Delta$ with multiplicities $4, \, 4$ and 8    are   $ 3, \, 2$ and $ {5\ov 2}$. This hints  at  % which is consistent with
     an effective 3d supersymmetry, but its     realization   for the  above   system of 8+8 scalars  and fermions  on $\AdS_{3,\b}$  should be non-trivial   as  it 
      appears to  require  the  the presence of the flat connection in  scalar and fermion covariant derivatives originating  from the twist in $\tilde S^4$.}
  
  Thus    we get from \rf{5.27},\rf{5.34}
  \ba \la{523}
  \G_1=\  &4\, \G^{(3,0)}(\beta)  + 2\,  \G^{(2,0)}(\beta)  + 2\,  \G^{(2,1)}(\beta)  -  8\,\G^{({5\ov 2} ,{1\ov 2} )}(\beta)\no \\
   = &\frac{\beta}{2}-\sum_{n=1}^{\infty}\frac{e^{-\b n}}{n} = \frac{\beta}{2}+\log(1-e^{-\beta})
= \log\Big(2 \sinh\frac{\beta}{2}\Big) \ . 
\ea
%A16
Like  in other    similar cases of supersymmetric M2  brane 1-loop effective actions
 we observe  remarkable  cancellations of all ``complicated''   contributions  that happen in the sum over  all fields.\foot{One
 %v3   corrr ????
   may draw an analogy  of  these  cancellations  with  what happens in the case of supersymmetric partition functions on $S^1 \times S^d$   that  are equivalent to superconformal indices  and thus  effectively  receive  contributions only from BPS states.
 Indeed, the prefactor  $ {1\ov 4 \sinh^2 {\beta\ov 2}}= {q \ov (1-q)^2} $  of the M2 brane instanton  contribution $e^{-N \beta}$ in \rf{5},\rf{2.8} that we reproduced as the 
  M2  brane partition function in \rf{425} 
 may be  also interpreted  \ci{Arai:2020uwd}   as the  superconformal index  of 
 $k=1$ abelian ABJM theory  \ci{Gang:2011xp}   or as a  supersymmetric partition function
 \ci{Closset:2013vra}  of 
a single  $\N=8$ 3d scalar supermultiplet in $S^1_\beta \times S^2$    background with extra 
 twist on $S^2$ required  for supersymmetry (corresponding to the   presence of rotation generator in the definition of the  3d  superconformal index). 
 Similar relation  may   apply  also  to the WL  computation in this section.
 }

The final  result   for the ``defect''  M2  brane 1-loop  partition function   is  very simple 
\be \la{524}
{\cal Z}_1= {1\ov 2 \sinh\frac{\beta}{2}}  \ , \ee 
and  thus  matches   the prefactor  in the  leading term in the Wilson loop expectation value  in (\ref{2.13}).

%%%%%%%%%%%%%%%%%%%%%%%%%%%%%%%
\section{Summary and concluding remarks}
\label{sec-concl}

Let   us summarize  what we have found above. 
We  considered  the semiclassical  expansion of the M2  brane partition function  $\rZ$  \rf{36}  in the  11d background  AdS$_{7,\b} \times \tS^4$ 
\rf{2},\rf{3} 
which is an $S^1_\b$-compactified and  ``twisted''   version of the  maximally supersymmetric  AdS$_{7} \times  S^4$   limit  \rf{1}
 of the multiple  M5 brane   solution of 11d supergravity. 
 The   main dimensionless  parameters  are $\b$ (the ratio  of the length of 11-circle   to the scale $a$ 
 of $\AdS_7$ in \rf{1},\rf{2})  and 
      the  effective M2 brane tension $\T_2$  (or $N$) 
       \be \la{60}
  \T_2 = a^3 T_2 = {2\ov \pi} N, \qquad \qquad    T_{2} = \frac{1}{(2\pi)^{2}\ell_{P}^{3}}, \ \ \ \qquad  
   a =  2 ( \pi N)^{1/3}  \ell_P \ . \ee 
 Our first example  was   the  ``instanton''  M2  brane solution that is wrapped on $S^1_\beta$ of   $\AdS_{7,\b}$   and $S^2$   of  $\tS^4$.
 We found that in this case (see \rf{41}-\rf{420},\rf{425}) 
  \be \la{61} 
S^1_\beta \times S^2\, : \qquad    \rZ  =    -\frac{1}{\big(2\sinh\frac{\beta}{2}\big)^2} \,  e^{-\T_2 \bar S_{\rm cl}}\big[1 + \OO (\T_2^{-1} )\big] \ , \ \ \ \ \ \qquad  
\te  \bar S_{\rm cl}=(1-\ha) \pi \b =  \ha \pi \b  \ . 
  \ee
We  also studied  the ``defect''  M2 brane solution   wrapped on the ``thermal''  $\AdS_{3,\b}$  part of $\AdS_{7,\b}$ %(and point-like in $\tS^4$) 
that  corresponds to   ``open'' M2  brane  ending on the $S^1\times S^1_\b$  at the boundary of  AdS$_{7,\b}$ 
%v3
  (thus  representing 
a Wilson-surface like  
``defect'' in the (2,0)  theory   that generalizes the circular BPS Wilson loop in gauge theory).\foot{The similar  AdS$_3$  ``defect'' 
M2  brane  solution  considered in \ci{Drukker:2020swu} has $S^2$  boundary instead of  $S^1\times S^1_\b$  and thus has logarithmically divergent 
 classical action related to the defect conformal  anomaly.}
In this case  (see \rf{52},\rf{524})
 \be \la{62} 
\AdS_{3,\b} \, : \qquad   \rZ =    \frac{1}{2\sinh \frac{\beta}{2}} \,  e^{-\T_2 \bar S_{\rm cl}}\big[1 + \OO (\T_2^{-1} )\big] \ , \qquad  \ \ \ \ \ 
\te  \bar S_{\rm cl}= -   \ha \pi \b  \ . 
  \ee
  It is useful   to compare  these results with what was found in 
  \ci{Giombi:2023vzu,Beccaria:2023ujc}     for   similar  M2 brane solutions  in 
  AdS$_4 \times S^7/\mathbb Z_k$  M-theory background  dual to $U_k(N) \times U_{-k}(N)$ 3d Chern-Simons-matter  ABJM theory 
   \ci{Aharony:2008ug}.  This  11d background is the  supersymmetric  $\ZZ_k$ orbifold 
 of the AdS$_4 \times S^7$   which is  a limit of  the multiple M2 brane solution of 11d supergravity  (cf. \rf{1},\rf{60}):
   \ba
\la{63}
 &\qquad \qquad \qquad ds^{2}_{11}= R^2\big(\tfrac{1}{4}ds^{2}_{\AdS_{4}}+ds^{2}_{S^{7}/\ZZ_k}\big), % \qquad
  \\
 & ds^{2}_{S^{7}/\ZZ_k} =ds^{2}_{\CP^{3}} +   (d\y +   A)^{2}\, , \qquad  \y \equiv  \y +  \rb \ , \qquad \  \rb \equiv{2 \pi \ov k} \ , \\
& F_4 = - \tfrac{3}{ 8}i  R^3 \vol_{\AdS_4} \ , \qquad \ \ R= \big(32\pi^{2}N k \big)^{1/6} \ell_P \ , \ \ \ \ \ \ \ \ 
\T_2 = R^3 T_2 =  {\sqrt{2k}\ov \pi } \sqrt N \ . \la{64}
 \ea
 We are assuming the Euclidean signature  and $A$   depends on the  6 coordinates of $\CP^3$. 
  Here the dimensionless  parameters are $k$ and $N$, or  $\rb$    and the effective   tension  $\T_2$. 
  
The  M2  brane ``instanton''    solution   considered in \ci{Beccaria:2023ujc}   is the 11d   uplift  of the IIA   string
 $\CP^1$ instanton of \ci{Cagnazzo:2009zh}: it  is wrapped on the 11d circle  $\y$  of   dimensionless   length    $\rb={2\pi\ov k}$ 
and on $\CP^1 \subset  \CP^3$,   so that
%A15
 it has the  $S^3/\ZZ_k$ world-volume  metric. 
In this case one finds   for the M2  brane partition function \ci{Beccaria:2023ujc}\foot{Here we
ignore the overall factor 4  that   accounts for contribution of  the  anti-instanton saddle 
and also for the effect of resolution of the 0-mode degeneracy (see \ci{Gautason:2023igo,Beccaria:2023ujc}).}
\be \la{65} 
S^3/\ZZ_k\, : \qquad    \rZ  =    \frac{1}{\big(2\sin \rb 
%  \frac{2\pi}{k}
\big)^2} \,  e^{-\T_2 \bar S_{\rm cl}}\big[1 + \OO (\T_2^{-1} )\big] \ ,  \ \qquad  
  \bar S_{\rm cl}= {\vol}(S^{3}/\ZZ_k) = \pi \rb  =  \tfrac{2\pi^2}{ k}  \ .
  \ee
This  corresponds to  the leading  $e^{-2\pi \sqrt{2N\ov k} }$   term in the  large $N$ 
non-perturbative part 
of the localization result  \ci{Drukker:2011zy,Hatsuda:2013gj}  for the free  energy of the ABJM theory on $S^3$. 

Another  M2 brane solution in \rf{63}   considered in \ci{Sakaguchi:2010dg,Giombi:2023vzu}
has  world-volume of $\AdS_2\times  S^1/\ZZ_k $    where   $S^1/\ZZ_k $   corresponds to the $\y$-circle in \rf{63}
and $\AdS_2 \subset \AdS_4$   has  the $S^1$ boundary. It
  may be  interpreted as  a dual  of the  circular  BPS  Wilson loop in the ABJM theory. 
In this case %one finds
 \ci{Giombi:2023vzu}
\be \la{66} 
\AdS_2\times  S^1/\ZZ_k\, : \quad    \rZ  =    \frac{1}{2\sin \rb } \,  e^{-\T_2 \bar S_{\rm cl}}\big[1 + \OO (\T_2^{-1} )\big]  ,   \quad  
  \bar S_{\rm cl}=\tfrac{1}{4}  {\vol}(\AdS_2) \rb = - \tfrac{1}{2}  \pi \rb  = -\tfrac{\pi^2}{ k}    \ .
  \ee
  This matches the leading large $N$ term $ [{2\sin \frac{2\pi}{k} }]^{-1} e^{\pi \sqrt{\frac{2N}{k}}}$ in the localization result \cite{Klemm:2012ii}  for the  $\ha$-BPS Wilson loop in the ABJM theory, in the limit of large $N$ with  $k$ fixed. 

Comparing \rf{61},\rf{62} with \rf{65},\rf{66} we observe close similarities. 
This suggests some relation by analytic continuation of both the backgrounds and the M2  brane solutions. 
Indeed, the  maximally supersymmetric AdS$_7 \times S^4$  and AdS$_4 \times S^7$   backgrounds are related by
 a formal analytic continuation  (like the one  between  AdS$_n$  and $S^n$, i.e.  $dx^2 + \sinh^2 x\ dS_{n-1}   \to - ( dr^2 + \sin^2 r\,  dS_{n-1} ), \  r= i x$)  and the same will apply to the M2  brane actions in these backgrounds. 
 %Then the  M2  brane actions in the two backgrounds   will be related by  changing the sign of the 
 
 The   compactification $y\equiv y + \b$  of the circle in $\AdS_{7,\b}$   part of  \rf{2}  suggests  an analogy with the  discrete 
  orbifolding  $\y \equiv \y + \rb$ in $S^7/\ZZ_k $ part of \rf{63}  and thus a similar role of $\b$  and $\rb$, 
  which is indeed   evident   from the  comparison of  \rf{61},\rf{62} with \rf{65},\rf{66}.
 %A15
  Such analytic continuation  suggests   that   %via this analytic continuation 
 the  ``instanton''   $S^1_\b \times S^2$   M2 solution in $\AdS_{7,\b}\times \tS^4$
 may   be related  to the ``Wilson loop'' $\AdS_2\times  S^1/\ZZ_k$   solution in $\AdS_{4}\times S^7/\ZZ_k$, 
 and 
vice versa, the ``defect'' $\AdS_{3,\b}$   solution in $\AdS_{7,\b}\times \tS^4$   may  be   related
 to the ``instanton''   $S^3/\ZZ_k$  solution in $\AdS_{4}\times S^7/\ZZ_k$.\foot{The factor  of 2  
 %A16
 mismatch  in powers of sinh/sin prefactors  in the  corresponding  M2 brane partition functions 
 may be related to the fact that  the analytic continuation
 maps a  world-volume with $S^1$ times a 2-sphere topology to  $S^1$ times a disk (AdS$_2$) one.}

 Still, some details do not match: the $\ZZ_k$ orbifold  of $S^7$ in  the Hopf fibration parametrization 
 is not equivalent to  an  analytic continuation  of  an orbifold of  $\AdS_7$   with $S^5\times S^1$   boundary.\foot{
The $S^7$ metric  can be parametrized    as 
$
Z^*_{r}Z_{r}=1 \ (r=1,2,3,4)
$ with 
$
Z_{r}=e^{i\y}W_{r}$ where  $\y = \y+2\pi$  and   $ W^*_r W_r =1$
 parametrize $\CP^{3}$ so  that (see, e.g., \cite{Nishioka:2008gz})
$
dS_7  = ds^{2}_{\CP^{3}}+(d\y+A)^{2}, 
$
where $A$ depends on $\CP^3$   coordinates. 
%where $w_{i}$ are three complex coordinates related to $W_{r}$, see  \eg Eq.~(2.6) in \cite{Nishioka:2008gz}.
Alternatively, we may set 
$Z_{1}=\cos r \, e^{iy}, \  Z_{i}=\sin r\, U_{i} \ ( i=1,2,3)$,   $  U_i^*U_{i}=1$ 
where $U_{i}$ parametrize $S^{5}$. Then  the  $S^7$  metric is 
$
dS_7 = dr^{2}+\sin^{2}r \, dS_{5}+\cosh^{2}r\, dy^{2}. 
$
To  relate this   to the first Hopf fibration parametrization 
  of the metric  we need  to redefine $U_i$  by   $e^{iy}$  and identify $y$ with $\y$. 
Then   the  orbifold   of  $\y$   will act also on $S^{5}$. But  orbifolding 
 $y\equiv y + \rm b $   in the second form of the   metric does  not act on $S^{5}$. 
Thus  the two   orbifolds are not equivalent.
% (at least in any  obvious ways). 
 }
 Also, there  is  no analog of the $z\to z+ i y$   twist in $\AdS_{7,\b}\times \tS^4$ in \rf{2}  on the 
 $\AdS_{4}\times S^7/\ZZ_k$  side. 
 Thus  the  reason for the 
 close  similarity   between the expressions in \rf{61},\rf{62}   and  \rf{65},\rf{66}  calls for further  insight.

\section*{Acknowledgments}

We are grateful to   J. Figueroa-O'Farrill,  V.G.M.  Puletti, J. Russo   and  K. Zarembo 
   for useful  discussions. 
   %A19
   We also thank 
   %v2
   P. Bomans,  F. Gautason, V.G.M.  Puletti  and J. van Muiden for helpful comments on the draft. 
MB was supported by the INFN grants GSS and GAST. 
SG is supported in part by the US NSF under Grant No. PHY-2209997. 
AAT is supported by the STFC grant ST/T000791/1.

%  \small

\appendix

\section{Renormalized volume of $\AdS_{2n+1}$ with boundary $S^{2n-1}\times S^1$}
\la{app:vol}

As is well  known, the   regularized  volume of global $\AdS_{2n+1}$  with $S^{2n}$ boundary has  logarithmic IR  divergence, 
$\vol(\AdS_{2n+1})=-  {2(-1)^n \pi^n \ov n!} \log \eps $ (where $\eps\to 0$);   in particular, 
$\vol(\AdS_{7})=  { \pi^3 \ov 3} \log \eps$  (see, e.g., \ci{Diaz:2007an}). 
At the same time, in the case of  $S^{2n-1}\times S^{1} $  boundary  the volume contains only power divergences and thus
 is finite after one drops them. This is 
 analogous to  the case of $\AdS_{2n}$  with $S^{2n-1}$ boundary where 
$\vol(\AdS_{2n})= {(-1)^n (2\pi)^n \ov (2n-1)!!}$. 

To find  the volume of $\AdS_{2n+1}$  with $ S^{2n-1} \times S^1$  boundary\foot{This space  may be viewed as
  ``thermal'' AdS$_{2n+1}$, i.e.  is obtained from  Minkowski signature $\AdS_{2n+1}$ by
analytic continuation and  periodical identification of  the Euclidean time.}
%Let us consider the regularization of volume of $AdS_{2n+1}$ with boundary $S^{1}\times S^{2n-1}$.
   %first consider the $AdS_{2n+1}$ metric in global coordinates  ($x\ge 0$)
\be\la{a1}
ds^2 = dx^2 +  \sinh^2 x\, dS_{2n-1} +  \cosh^2 x \,dy^2 \ , \ \ \qquad \ \   y\equiv y+2\pi  \ . 
\ee
%with boundary $R\times S^{2n-1}$. Setting $t=iy$ and $y\equiv y+2\pi$, this is $S^{1}\times S^{2n-1}$. Volume with 
Let us introduce an IR   cutoff $0<x\le x_{0}$  in the volume integral 
\be
\la{C.2}
\vol(\AdS_{2n+1}) =  \text{vol}(S^{2n-1}\times S^1)\ \int_0^{x_{0}} dx\, \cosh x\, \sinh^{2n-1} x =
   \text{vol}( S^{2n-1}\times S^1)\ \frac{1}{2n}\sinh^{2n}x_{0} \, .
\ee
A  natural cutoff is $r=  \eps^2 \to 0 $ in Fefferman-Graham coordinates 
$
ds^2 = \frac{1}{4r^{2}} dr^2   + \frac{1}{r}  g_{mn}(x,r)  dx^m dx^n $
which is related to $x_0$ as $x_{0}=- \log\eps$.\foot{For comparison, in the case  of  $\AdS_{2n+1}$ with $S^{2n}$ boundary, i.e.  $ds^{2} = dx^{2}+\sinh^{2}x\, dS_{2n}$, we get 
$
\int_{0}^{x_{0}}dx\, \sinh^{2n}x = \frac{(-1)^{n}\Gamma(n+\frac{1}{2})}{\sqrt\pi\Gamma(n+1)}\, x_{0} + \cdots,
$
where dots  stand for  powers of $\sinh x_{0}$  leading  to    powers of $1\ov \eps$ and subleading 
finite terms.    Multiplying by $\text{vol}(S^{2n})$, one gets
$\vol(\AdS_{2n+1}) = - \frac{2(-1)^{n}\pi^{n}}{n!}\, \log \eps$.} 
Dropping $1\ov \eps^k$ power divergences in \rf{C.2}  and setting $\eps\to 0$  gives
(using that $\text{vol}(S^{n}) = \frac{2\pi^{\frac{n+1}{2}}}{\Gamma(\frac{n+1}{2})}$)  
\be
\vol(\AdS_{2n+1}) = 
  \vol(S^{2n-1}\times S^1 )\ \frac{(1-\eps^2)^{2n}}{2^{2n+1}\,n\,\eps^{2n}} 
  \ \to \   \text{vol}(S^{2n-1}\times S^1 ) %\,\Big[\text{power div}
\frac{(-1)^{n}\Gamma(n+\frac{1}{2})}{2n^{2}\sqrt\pi\Gamma(n)} 
=  \frac{(-1)^{n}\pi^{n+1}(2n)!}{2^{2n-1}\,(n!)^{3}}.\la{a2}
\ee
In particular,
  \be\la{a3}  \vol(\AdS_{3})=-\pi^{2}, \qquad 
 \vol(\AdS_{5})=  \frac{3\pi^{3}}{8}, \qquad  \vol(\AdS_{7}) = -\frac{5\pi^{4}}{48}\ . \ee
%\frac{35\pi^{5}}{1536}$.
As an application, let  compute the  value of the 11d supergravity action on 
$\AdS_{7,\b}\times S^4 $  of radius $a$  where $\b$ is the length of the   $S^1$ circle $y$  as in \rf{2}. 
Compactifying on $S^4$  (that has  radius $a\ov 2$)  we get 
%The bulk action is 
\be\la{a4}
S_{11}  = -\frac{1 }{2\kappa_{11}^{2}}\,  ({a\ov 2})^4 \vol(S^{4})\, \int d^{7}x \sqrt{g}\, \big(R^{(7)}-2\Lambda\big), \qquad
\qquad  2\kappa_{11}^{2}
= (2\pi)^{8}\ell_{P}^{9}  \ ,    %= \frac{\pi^{5}a^{9}}{2N^{3}}.
\ee
where  for  the AdS$_7$ solution one has  $R^{(7)}= -{42\ov a^2}$,   $\Lambda= -{15\ov a^2}$. %  and $a$ is the radius of $\AdS_7$ as in \rf{1}.  
%From the equations of motion  \be
%R_{mn}-\frac{1}{2}g_{mn}(R^{(7)}-2\Lambda) = 0 \  \to \mc R-2\Lambda = \frac{2}{7}\mc R = \frac{2}{7}\frac{-6\cdot 7}{a^{2}}= -\frac{12}{a^{2}}.
%\ee
Since   $\vol(S^{4}) = \frac{8\pi^{2}}{3}$, we get
\be
S_{11} = -  {1\ov (2\pi)^8} ({a\ov \ell_P})^9  \frac{8\pi^{2}}{3}(\ha )^{4}(-12)\,{\beta\ov 2\pi} 
\vol(\AdS_{7}) = {\pi \ov (2\pi)^8} ({a\ov \ell_P})^9  \beta\,\vol(\AdS_{7}) \ . \la{a5}
\ee
Using  \rf{a3}, i.e.   $ \vol(\AdS_{7}) = -\frac{5\pi^{4}}{48}$, and  \rf{1}   implying 
$({a\ov \ell_P})^9 = 2^9 \pi^3 N^3 $  we end up with 
\be 
S_{11}=   -\frac{5}{24}N^{3}\beta \ . \la{a6}  \ee
The same result is found also  for the ``twisted'' $\AdS_{7,\b}\times \tS^4 $  background in \rf{2} 
(the shift  $z \to z + i y$   along the $S^4$ isometry $z$-direction 
 does not change the value of the 11d   volume   form $\sim dy \wedge dz\wedge ...$). 
 At the same time, the leading  large $N$  term in the   free energy in \rf{29}, 
is $F_N^{\rm pert}= - {1\ov 6} N^3\b  + ..., $  so there is a $5/4$ mismatch  with \rf{a6}. 

%v2
 This  discrepancy  was  noted   in \cite{Kallen:2012zn,Minahan:2013jwa}, 
see also  \cite{Minahan:2016xwk}. 
A way to resolve it  at the level of 7d gauged supergravity with extra (non-invariant) counterterms 
was suggested in \ci{Bobev:2019bvq}. 
%A priori, it  may seem   there is no  % reason why  one should expect an exact  match as this 
It is unclear at the moment how  to reach the same conclusion  directly at the level of 11d supergravity action, i.e. to see how the 
   leading-order term  can   distinguish between  the 
 standard and  ``supersymmetric'' free  energy.
 %v2
One may contemplate adding  some   non-invariant boundary terms, 
but %this seems a priori problematic and
 this  issue needs further clarification.

\section{$\beta\to 0$ expansion of scalar free energy in thermal $\AdS_{3,\b}$} % of thermal determinants}
\la{app:smallbeta}

Here  we discuss several methods to compute the small $\beta$ expansion of the non-Casimir part of the scalar  log det in   (\ref{5.27}), i.e. of the function 
\be
\la{D.1}
f(\beta; \Delta) \equiv  \sum_{n=1}^{\infty}\frac{q^{n\Delta }}{n(1-q^{n})^{2}},\qquad \ \ \    q= e^{-\b q} \ , \qquad  \Delta \ge 2,
\ee
that can be written equivalently as 
\be
\la{D.2}
f(\beta; \Delta)  = -\sum_{\ell,\ell'=0}^{\infty}\log(1-q^{\ell+\ell'+\Delta}) = 
 -\sum_{n=0}^{\infty}(n+1)\log(1-q^{n+\Delta}).
\ee
%\paragraph{Method I}
The  first method  is to expand $f$ in  (\ref{D.1}) at small $\beta$ and sum the terms using Riemann   zeta-function regularization 
 (i.e. multiplying by $n^{s}$, summing, and taking the finite part of the $s\to 0$ limit).
This gives 
\ba
\la{D.3}
\text{(I):} \qquad  f(\beta; \Delta) = &\frac{\zeta (3)}{\beta ^2}-\frac{\pi ^2 (-1+\Delta )}{6 \beta 
}+\frac{1}{12} \gamma_{\rm E}  (5-12 \Delta +6 \Delta ^2)+\frac{1}{24} 
(-1+\Delta ) (1-4 \Delta +2 \Delta ^2) \beta \lp
+\frac{(-1+20 \Delta -50 
\Delta ^2+40 \Delta ^3-10 \Delta ^4) }{2880}\beta ^2\lp
+\frac{(-5+42 
\Delta +63 \Delta ^2-420 \Delta ^3+525 \Delta ^4-252 \Delta ^5+42 
\Delta ^6) }{3628800}\beta ^4+\cdots. 
\ea
The constant $\g_{\rm E}$ term  is regularization dependent and  is related to  the dropped pole $\sim {1\ov s}$.

%\paragraph{Method II}

Another   method is to expand $f$ in \rf{D.2} at small $\beta$, multiply by $(n+\Delta)^{s}$,
sum over $n$  and then take the finite part of the $s\to 0$ limit.
This way we obtain 
\ba
\la{D.4}
\text{(II):} \qquad  f(\beta; \Delta) =&\mathscr C(\Delta)
-\frac{1}{12} (5-12 \Delta +6 \Delta ^2) \log  \beta +\frac{1}{24} 
(-1+\Delta ) (1-4 \Delta +2 \Delta ^2) \beta \lp
+\frac{(-1+20 \Delta -50 
\Delta ^2+40 \Delta ^3-10 \Delta ^4) }{2880}\beta ^2 \lp
+\frac{(-5+42 
\Delta +63 \Delta ^2-420 \Delta ^3+525 \Delta ^4-252 \Delta ^5+42 
\Delta ^6) }{3628800}\beta ^4+\cdots\ , \\
\la{D.5}
\mathscr C(\Delta) = & (\Delta-1 )\Big [\frac{1}{2} \log (2 \pi )-\log\Gamma(\Delta )\Big]+\zeta'(-1,\Delta ).
\ea
Comparing to  (\ref{D.3}), we see that we miss the $1\ov \beta^{2}$ and $1\ov \beta$ terms and the $\gamma_{\rm E}$ term is
 replaced by the   $\log\beta$ term.

%\paragraph{Method III}

A   rigorous (third) method is to   follow   \cite{Cardy:1991kr}.\footnote{Another rigorous approach is based on the temperature inversion relations as in \cite{Gibbons:2006ij}.
}
Starting again from (\ref{D.2}) and differentiating over $\b$  gives 
\be
f'(\beta; \Delta) =-\sum_{n=0}^{\infty}\frac{(n+1)(n+\Delta)}{e^{(\Delta+n)\beta}-1} = -\sum_{n=0}^{\infty}\sum_{m=1}^{\infty}(n+1)(n+\Delta)e^{-(\Delta+n)m\beta} .
\ee
Now using that  
$
e^{-x} = \frac{1}{2\pi i}\int_{C}ds\, x^{-s}\,\Gamma(s)
$  (where the contour $C$ is along the imaginary axis with large enough real part of $s$)
 gives the Mellin representation
\ba
f'(\beta; \Delta) &= \frac{1}{2\pi i}\int_{C}ds \sum_{n=0}^{\infty}\sum_{m=1}^{\infty}(\Delta+n)^{-s}m^{-s}\beta^{-s}\Gamma(s)(n+1)(n+\Delta)
= \frac{1}{2\pi i}\int_{C}ds\, \beta^{-s}G(s),
\\
G(s) & = \Gamma(s)[\zeta(-2+s,\Delta)+(1-\Delta)\zeta(-1+s,\Delta)]\zeta(s).
\ea
Closing the contour to the left we get  for the $\b \to 0$ expansion (up to exponentially suppressed terms denoted by dots)
\ba
f'(\beta; \Delta) &= -\sum_{n=0}^{\infty}\mathop{\text{Res}}_{s=3-n}(\beta^{-s}G(s))+... %\text{non-perturbative}.
\ea
Integrating this over  $\beta$  gives   %(we omit the non-perturbative part)
\ba
\la{D.11}
\text{(III):} \qquad  f(\beta; \Delta) =&\mathscr C(\Delta)+\frac{\zeta (3)}{\beta ^2}-\frac{\pi ^2 (\Delta-1 )}{6 \beta 
}-\frac{1}{12} (5-12 \Delta +6 \Delta ^2) \log \beta \lp
+\frac{1}{24} 
(-1+\Delta ) (1-4 \Delta +2 \Delta ^2) \beta +\frac{(-1+20 \Delta -50 
\Delta ^2+40 \Delta ^3-10 \Delta ^4)}{2880} \beta ^2\lp
+\frac{(-5+42 
\Delta +63 \Delta ^2-420 \Delta ^3+525 \Delta ^4-252 \Delta ^5+42 
\Delta ^6) }{3628800}\beta ^4 +\cdots, 
\ea
where $\mathscr C (\Delta)$ is  yet undetermined   integration constant. 
By doing numerics, we found that (\ref{D.11}) is the correct expansion with %, \cf (\ref{D.5}),
$\mathscr C(\Delta)$  being the same as  in \rf{D.5}. 
The expansion \rf{D.11}  reproduces 
 the two singular $1\ov \b^2$  and $1\ov \b$ terms in (\ref{D.3}) and the logarithm in (\ref{D.4}).

For example, this gives for $\Delta=3$
\ba\la{b11}
f(\beta; 3) &= \frac{\zeta (3)}{\beta ^2}-\frac{\pi ^2}{3 \beta }+\frac{1}{12}\Big(1 -12 
\log \frac{\mathsf A}{2 \pi }\Big)-\frac{23}{12} \log \beta +\frac{7 \beta 
}{12}-\frac{121 \beta ^2}{2880}+\frac{251 \beta ^4}{725760} %-\frac{241 \beta ^6}{43545600}
+\cdots.
\ea

\section{Scalar determinant  in  $\AdS_{3,\b}$  from   expansion in modes on $S^{1}\times S^{1}_{\beta}$}
\la{app:s1s1}

Here  we derive the  expression in \rf{5.27}
by  directly expanding in Fourier modes  in  the two   $S^{1}\times S^{1}_{\beta}$   boundary angles.\foot{As usual, 
the determinant will be defined   using  analytic  regularization  so   that power divergences   will be  ignored  
(there  is no logarithmic divergence in the present 3d  case).}
   % thermal functional determinant for a scalar field by direct 
%reduction as a sum over modes along the two boundary circles. To this aim, 
 Let us  start with    the scalar  operator $\wh K\equiv \Delta_0 = - D^2 + M^2$  in the $\AdS_{3,\b}$ metric  \rf{51} 
 in  the  explicit coordinate form  (here $M^2= \Delta(\Delta-2)$ as in \rf{514}) 
\ba
%S &= \frac{1}{2}\int d^{3}\xi\, \sqrt{G}\, z \wh K z, \qquad \sqrt{G}=\sinh\xi_{1}\cosh\xi_{1},\\
\wh K  &= -\frac{1}{\sinh\xi_{1}\cosh\xi_{1}}\partial_{1}(\sinh\xi_{1}\cosh\xi_{1}\partial_{1})-\frac{1}{\sinh^{2}\xi_{1}}\partial^{2}_{2}
-\frac{1}{\cosh^{2}\xi_{1}}\partial_{3}^{2}+\Delta(\Delta-2)\ .
\ea
Redefining  $\xi_{1}\to \frac{1}{2}\rho$ and  expanding in modes %replace 
so that $
\partial_{2} \to i\,m, \  \partial_{3}\to i\,n_{_{\beta}} = i\, \frac{2\pi}{\beta}n,
$
 we get  a  ``radial''  1d operator  % \footnote{Notice that the operator is invariant under $\Delta\to 2-\Delta$.}
%v3
\be
K_{m,n} = -\frac{4}{\sinh\rho}\frac{d}{d\rho}\Big(\sinh\rho\,\frac{d}{d\rho}\Big)+\frac{m^{2}}{\sinh^{2}\frac{\rho}{2}}
+\frac{n_{_{\beta}}^{2}}{\cosh^{2}\frac{\rho}{2}}+\Delta(\Delta-2), \qquad n_{_{\beta}} = 
 \frac{2\pi}{\beta}n\ , \quad  n,m\in \Z.
\ee
By applying the Gelfand-Yaglom theorem (see, e.g.,
%v3
  \cite{Dunne:2007rt,Kruczenski:2008zk,Aguilera-Damia:2018rjb})  we have
\ba\la{c3} 
\log\frac{\det K_{m,n}}{\det K_{m,0}} =  &\lim_{\rho\to \infty}\log\frac{\psi_{m,n}(\rho)}{\psi_{m,0}(\rho)},
\\
 K_{m,n}\, \psi_{m,n}(\rho) = & 0, \qquad \psi_{m,n}(\rho) \stackrel{\rho\to 0}{\to}\rho^{|m|}+\cdots. \la{c4}
\ea
%v3  ????  corr  
The solution  of \rf{c4} is 
\be
\psi_{m,n}(\rho) = 2^{|m|}(\tanh\frac{\rho}{2})^{|m|}\,(\cosh\frac{\rho}{2})^{-\Delta}\ {}_{2}F_{1}
\Big(\frac{\Delta+|m|-i n_{_{\beta}}}{2}, \frac{\Delta+|m|+in_{_{\beta}}}{2}, 1+|m|, \tanh^{2}\frac{\rho}{2}\Big),
\ee
and  as a consequence of \rf{c3} 
%v3
\be
\la{E.8}
\log\frac{\det K_{m,n}}{\det K_{m,0}} = \log\frac{\Gamma(\frac{\Delta}{2}+\frac{|m|}{2})^{2}}
{\Gamma(\frac{\Delta}{2}+\frac{|m|}{2}-i\frac{n_{_{\beta}}}{2})\,\Gamma(\frac{\Delta}{2}+\frac{|m|}{2}+i\frac{n_{_{\beta}}}{2})}.
\ee
Thus  % (exploiting $\sum_{n\in\Z}1 = 2\zeta(0)+1=0$ to drop the $n$ independent terms)
\be
\la{E.9}
\Gamma^{(\Delta)} (\b) \equiv  
\ha \log \det \wh K =  \frac{1}{2}\sum_{n,m\in \Z}\log\det K_{m,n} = - \frac{1}{2}\sum_{n,m\in \Z}\log\Big[
\Gamma\Big(\frac{\Delta}{2}+\frac{|m|}{2}-i\frac{n_{_{\beta}}}{2}\Big)\,\Gamma\Big(\frac{\Delta}{2}+\frac{|m|}{2}+i\frac{n_{_{\beta}}}{2}\Big)\Big], 
\ee
where  we dropped $n$-independent  term  as $\sum_{n\in\Z}1 = 1 + 2\zeta_R(0)=0$.
As in   \cite{Denef:2009kn}  we   may use that 
\be
\log\big[\Gamma(x+iy)\Gamma(x-iy)\big] = 2\log\Gamma(x)-\sum_{k=0}^{\infty}\log\Big[1+\frac{y^{2}}{(x+k)^{2}}\Big].
\ee
Then from  (\ref{E.9})  we get %($x=\frac{\Delta}{2}+\frac{|m|}{2}$ does not depend on $n$)
\ba
\Gamma^{(\Delta)}(\b) &= \frac{1}{2}\sum_{n,m\in \Z}\sum_{k=0}^{\infty}\log\Big[1+\frac{n_{_{\beta}}^{2}}{(\Delta+|m|+2k)^{2}}\Big].
\ea
The set $|m|+2k$ with $m\in\Z$ and $k\in \mathbb{N}_{0}$ can be replaced by a sum over $k\in \mathbb{N}_{0}$ with multiplicity $k+1$. 
Thus,
\ba
\Gamma^{(\Delta)}(\b) &= \frac{1}{2}\sum_{n\in \Z}\sum_{k=0}^{\infty}(k+1)\log\Big[1+\frac{n_{_{\beta}}^{2}}{(\Delta+k)^{2}}\Big].
\ea
The  $n=0$ term  vanishes and separating 
%v3
 the divergent part  of the sum over $n$ we get\foot{Note
 %v3
   that  here the  "reg" part may still contain 
 a divergent contribution from the sum over $k$ (see below).}
\ba
&\qquad \qquad \qquad  \Gamma^{(\Delta)}(\b) =\Gamma_{\rm div}^{(\Delta)}(\b) +\Gamma_{\rm reg}^{(\Delta)}(\b)\ ,\la{c11}
\\ 
\Gamma_{\rm div} ^{(\Delta)}(\b)&= \sum_{n=1}^{\infty}\sum_{k=0}^{\infty}(k+1)\log\frac{n_{_{\beta}}^{2}}{(\Delta+k)^{2}}, \qquad 
\Gamma_{\rm reg} ^{(\Delta)}(\b)= \sum_{n=1}^{\infty}\sum_{k=0}^{\infty}(k+1)\log\Big[1+\frac{(\Delta+k)^{2}}{n_{_{\beta}}^{2}}\Big].\la{c12}
\ea
%In the 
Computing $\Gamma_{\rm div}^{(\Delta)}(\b)$    using  again the Riemann zeta-function regularization gives 
%divergent piece we sum over $n$ with simple zeta function with $n^{s}$. This gives
\ba
\la{E.16}
&\Gamma_{\rm div} ^{(\Delta)}(\b)= \sum_{k=0}^{\infty}(k+1)\sum_{n=1}^{\infty}\log\frac{n_{_{\beta}}^{2}}{(\Delta+k)^{2}} 
= \sum_{k=0}^{\infty}(k+1)\sum_{n=1}^{\infty}[-2\log(\Delta+k)+2\log\frac{2\pi}{\beta}+2\log n] \no \\
&\qquad \quad= \sum_{k=0}^{\infty}(k+1)\big[\log(\Delta+k)-\log\frac{2\pi}{\beta}+\log(2\pi)\big] 
= \sum_{k=0}^{\infty}(k+1)\big[\log(\Delta+k)+\log\beta\big]\ . 
\ea
%v3
Here the   sum over $k$  may also  be  computed  using zeta-function regularization  regularization but 
it is useful   not to do this before  combining it with $\Gamma_{\rm reg} ^{(\Delta)}(\b)$. 

Since 
\be\la{c14}
\sum_{n=1}^{\infty}\log\Big(1+\frac{a^{2}}{n^{2}}\Big) = \log\frac{\sinh(\pi a)}{\pi a} = \pi a-\log(\pi a)-\log 2+\log(1-e^{-2\pi a})\ ,
\ee
 we find that $\Gamma_{\rm reg}^{(\Delta)}(\b)$ in  \rf{c12}
 %v3
  (here for $a = \frac{1}{2\pi}(\Delta+k)\beta$)
may be written as 
\ba
\Gamma_{\rm reg}^{(\Delta)}(\b) %&=\sum_{k=0}^{\infty}(k+1)\Big[\pi\frac{1}{2\pi}(\Delta+k)\beta-\log(\pi \frac{1}{2\pi}(\Delta+k)\beta)
%-\log 2+\log\Big(1-e^{-(\Delta+k)\beta}\Big)\Big]\lp
=\sum_{k=0}^{\infty}(k+1)\Big[\frac{1}{2}(\Delta+k)\beta-\log((\Delta+k)\beta)+\log(1-e^{-(\Delta+k)\beta})\Big].  \la{c15}
\ea
Adding  (\ref{E.16})  and  \rf{c15}  gives 
\ba
\Gamma^{(\Delta)}(\b) &
%\sum_{k=0}^{\infty}(k+1)\Big[\frac{1}{2}(\Delta+k)\beta+\log(1-e^{-(\Delta+k)\beta})\Big] \lp
= \tfrac{1}{2}\b \sum_{k=0}^{\infty}(k+1)(\Delta+k)\, + \sum_{k=0}^{\infty}(k+1)\log\Big(1-e^{- (\Delta+k)\beta}\Big).
\ea
Doing the sum in the first term using  Hurwitz  zeta-function  regularization  gives finally
the  expression  \cite{Giombi:2008vd}  equivalent  (cf. \rf{D.1},\rf{D.2}) 
to the one  in \rf{5.27},\rf{515}
\ba
\la{E.20}
\Gamma^{(\Delta)}(\b) & 
%-\sum_{k=0}^{\infty}(k+1)[\frac{1}{2}(\Delta+k)\beta+\log(1-e^{-2\pi \frac{1}{2\pi}(\Delta+k)\beta})] \lp
= \frac{1}{24}(\Delta-1)(1-4\Delta+2\Delta^{2})\,\beta +   \sum_{k=0}^{\infty}(k+1)\log(1-q^{\Delta+k}) \ . 
\ea
 Note that  in  the $\b\to 0$ expansion  the first ``Casimir'' term  
  cancels against the linear in $\b$ term  in the second term in \rf{E.20}  (see  (\ref{D.11})).

%Notice that the linear in $\beta$ term cancels with the corresponding term in (\ref{D.11})
% in the high temperature expansion.

\subsection{Including  the twist    $\del_3 \to \del_3 - \k$   or  $\nb \to \nb + i \k$ }
%$S^{1}\times S^{1}_{\beta}$ reduction with $i\kappa$ shift in $n_{_{\beta}}$}

Let us  now consider the determinant  of the scalar operator  including  the coupling to the flat  gauge potential in the
$\x^3$ direction \rf{526},\rf{con}, i.e. $\del_3 \to \del_3 -  \k$ or $\nb \to \nb + i \k $ where $\nb = {\b \ov 2 \pi}$. 
Repeating the  above  calculation with $n_{_{\beta}}\to n_{_{\beta}}+i\kappa$
we get in \rf{c14}  % amounts to the following replacement in the above derivation
\be
\sum_{n\in\Z} \log\Big(1+\frac{a^{2}}{n^{2}}\Big) \to 
\sum_{n\in\Z} \log\Big(1+\frac{a^{2}}{(n+i\frac{\beta\kappa}{2\pi})^{2}}\Big),
\ee
where the sum   can  be computed using 
%\footnote{In the l.h.s. we implicitly combine terms with $n$ and $-n$ into a real single logarithm.}
\be
\sum_{n\in\Z} \log\Big(1+\frac{a^{2}}{(n+ib)^{2}}\Big)  = \log\Big|1-\frac{\sinh^{2}(\pi a)}{\sinh^{2}(\pi b)}\Big|  \ . 
\ee
This leads to the following  modification of the expression \rf{c11},\rf{c12} for the  determinant in \rf{E.9} 
\ba
\la{E.23} \Gamma^{(\Delta,\k)}(\b) =\Gamma^{(\Delta,\k)}_{\rm div}(\b) +\Gamma^{(\Delta,\k)}_{\rm reg}(\b)\ , &\qquad 
\Gamma^{(\Delta,\kappa)}_{\rm div}(\beta) = \frac{1}{2}\sum_{n\in\Z}^{\infty}\sum_{k=0}^{\infty}(k+1)\log\frac{(n_{_{\beta}}+i\kappa)^{2}}{(\Delta+k)^{2}},\\
\Gamma^{(\Delta,\kappa)}_{\rm reg}(\beta) &= \frac{1}{2}\sum_{k=0}^{\infty}(k+1)
\log\Big[\frac{\sinh^{2}(\frac{(k+\Delta)\beta}{2})}{\sinh^{2}(\frac{\beta\kappa}{2})}-1\Big].
\ea
This can be written in a form similar to (\ref{D.1}) as follows. 
%v3
For  the divergent part  of the sum over $n$ 
 we get (ignoring  again a sum of a constant  assuming  $\zeta_R$ regularization) 
\ba
&\Gamma^{(\Delta,\kappa)}_{\rm div}(\beta) 
 %-\frac{1}{2}\sum_{k=0}^{\infty}(k+1)\sum_{n\in \Z}^{\infty}\Big[-2\log(\Delta+k)+2\log\frac{2\pi}{\beta}+\log[ (n+i\frac{\beta\kappa}{2\pi})^{2}]\Big] \lp
= \frac{1}{2}\sum_{k=0}^{\infty}(k+1)\Big[\log(\frac{\beta^{2}\kappa^{2}}{4\pi^{2}})
+\sum_{n=1}^{\infty}\log\big[ (n+i\frac{\beta\kappa}{2\pi})^{2}\big]
+\sum_{n=1}^{\infty}\log\big[ (n-i\frac{\beta\kappa}{2\pi})^{2}\big]\Big] \lp
= \frac{1}{2}\sum_{k=0}^{\infty}(k+1)\Big[\log(\frac{\beta^{2}\kappa^{2}}{4\pi^{2}})+2\log\big(
\frac{4\pi}{\beta\kappa}\sinh\frac{\beta \kappa}{2}\big)\Big]
%v3
= \sum_{k=0}^{\infty}(k+1) \ \log(2\sinh\frac{\beta \kappa}{2}) \ . \la{c27}
\ea
%v3
%up to a term $\log(-1)$ that is clearly a spurious part of the summation. \red{\Large [a better explanation/comment ??]}
 Using  that 
\ba
\log& \Big[\frac{\sinh^{2}\frac{(k+\Delta)\beta}{2}}{\sinh^{2}\frac{\k \beta}{2}}-1\Big]
% = -2\log(2\sinh\frac{\beta \kappa}{2})+
%\log\Big[4\sinh^{2}(\frac{(k+\Delta)\beta}{2})-4\sinh^{2}(\frac{\beta\kappa}{2})\Big] \lp
= -2\log(2\sinh\tfrac{\beta \kappa}{2})+\beta(k+\Delta)+\log[(1-q^{k+\Delta+\kappa})(1-q^{k+\Delta-\kappa})] \ , \la{c28}
\ea
for the $\Gamma^{(\Delta,\kappa)}_{\rm reg}(\beta) $  part   we get\footnote{The Casimir term is computed by splitting $\Delta =\frac{1}{2}(\Delta+\kappa)+\frac{1}{2}(\Delta-\kappa)$ and  using Hurwitz  zeta function regularization, i.e.  introducing a factor  $(k+\Delta\pm\kappa)^{s}$  and dropping singular terms in the limit $s\to 0$.}
\ba
\la{E.27}
&\Gamma^{(\Delta,\kappa)}_{\rm reg}(\beta) = \frac{1}{2}\sum_{k=0}^{\infty}(k+1)\Big[-2\log(2\sinh\tfrac{\beta \kappa}{2})+\beta(k+\Delta)
+\sum_{\pm}\log(1-q^{k+\Delta\pm \kappa})\Big] \\\
%MB1 removed 1/2 in first term in second and third line
&=   -\sum_{k=0}^{\infty}(k+1)\, \log(2\sinh\tfrac{\beta \kappa}{2})+\tfrac{1}{24}(\Delta-1)(1-4\Delta+2\Delta^{2}+6\kappa^{2})\beta
+ \tfrac{1}{2}\sum_{\pm}\sum_{\ell,\ell'=0}^{\infty}\log(1-q^{\ell+\ell'+\Delta\pm \kappa}) \lp
=     - \sum_{k=0}^{\infty}(k+1)\, \log(2\sinh\frac{\beta \kappa}{2})
+\tfrac{1}{24}(\Delta-1)(1-4\Delta+2\Delta^{2}+6\kappa^{2})\beta
-\tfrac{1}{2}\sum_{\pm}\sum_{n=1}^{\infty}\frac{1}{n}\frac{q^{n(\Delta\pm \kappa})}{(1-q^{n})^{2}}.\no 
\ea
Adding together \rf{c27}  and \rf{E.27}   we finally   get the  finite  expression  quoted  in \rf{5.34},\rf{519} 
\ba
\la{E.28}
\Gamma^{(\Delta,\kappa)}(\beta) &= 
 \frac{1}{24}(\Delta-1)(1-4\Delta+2\Delta^{2}+6\kappa^{2})\beta
-\frac{1}{2}\sum_{n=1}^{\infty}\frac{1}{n}\frac{q^{n(\Delta+ \kappa})}{(1-q^{n})^{2}} 
-\frac{1}{2}\sum_{n=1}^{\infty}\frac{1}{n}\frac{q^{n(\Delta- \kappa})}{(1-q^{n})^{2}}  \ . 
\ea
%\ie a Casimir contribution plus a simple modification of the $\kappa=0$ thermal determinant.
The  small $\b$ expansion of $\Gamma^{(\Delta,\kappa)}(\beta) $   can be  found as in \rf{D.1},\rf{D.11}:
%Expanding this with Method I -- assuming as for $\kappa=0$ that the coefficient of $\log(\beta)$ is same as minus the %coefficient 
%of $\gamma_{\rm E}$ -- one finally gets (the Casimir contribution cancels as for $\kappa=0$)
\ba\la{c31} \Gamma^{(\Delta,\kappa)}(\beta) =
- \frac{\zeta (3)}{\beta ^2}-\frac{\pi ^2 (\Delta-1)}{6 \beta }
- \mc C(\Delta, \kappa)
+\frac{1}{12}(5-12\Delta+6\Delta^{2}+6\kappa^{2})\,\log\beta
- \sum_{n=1}^{\infty} \mc C_{2n}(\Delta, \kappa)\,\beta^{2n},
\ea
where 
\ba
&  \mc C(\Delta, \kappa)= \ha [  \mc C(\Delta+ \kappa)  + \mc C(\Delta- \kappa)  ] \ , \la{c300}\\
&\mc C_{2}(\Delta, \kappa) = \frac{-1+20 \Delta -50 \Delta ^2+40 \Delta ^3-10 \Delta 
^4}{2880}+\frac{1}{288} (-5+12 \Delta -6 \Delta ^2) \kappa 
^2-\frac{\kappa ^4}{288}, \\
%%%%
&\mc C_{4}(\Delta, \kappa) =\frac{-5+42 \Delta +63 \Delta ^2-420 \Delta ^3+525 \Delta ^4-252 
\Delta ^5+42 \Delta ^6}{3628800}\lp\qquad \qquad \quad
+\frac{(1-20 \Delta +50 \Delta ^2-40 
\Delta ^3+10 \Delta ^4) \kappa ^2}{57600}+\frac{(5-12 \Delta +6 
\Delta ^2) \kappa ^4}{34560}+\frac{\kappa ^6}{86400}\ , \ \  ...
\ea

%This is the  thermal determinant in  \cite{Giombi:2008vd} including the Casimir term (that was omitted there).  
\subsection{Alternative derivation by Poisson resummation}

An alternative way to  derive the expression for the log det in \rf{E.9}   is to  apply the 
Poisson resummation  trick 
\be
\sum_{n\in\Z}f(n) = \sum_{\ell\in\Z}\wt f(\ell), \qquad\qquad   \wt f(\ell) = \mc F[f] \equiv  \int_{-\infty}^{\infty}dn\, f(n)\, e^{-2\pi i \ell n}.
\ee
Since 
\be
\mc F\big[\log(1+a^{2}n^{2})\big] = -\frac{1}{|\ell|}\exp\Big(-\frac{2\pi|\ell|}{|a|}\Big),
\ee
this gives
\ba
\Gamma^{(\Delta)}(\b) &= \frac{1}{2}\sum_{\ell\in \Z}\sum_{k=0}^{\infty}(k+1)\frac{1}{|\ell|}e^{-|\ell|(\Delta+k)\beta}\ . 
\ea
If we separate  the $\ell=0$ term, we obtain 
\ba
\Gamma^{(\Delta)}(\b) &= ``\ell=0\ \text{term}''  + \sum_{k=0}^{\infty}(k+1)\log(1-e^{-(\Delta+k)\beta}).
\ea
This   can be generalized  to the case of  a non-zero $\k$-shift using that 
% Assuming one can do a complex shift in the FT variable $n\to n-ib$ one gets
\be
\mc F\big[\log(1+a^{2}(n+i b)^{2}\big] = e^{2\pi \ell b}\ \mc F[\log(1+a^{2}n^{2})],
\ee
which  leads to  the last term in (\ref{E.27}).

\iffa 
%%%%
\mc C_{6}(\Delta, \kappa) =& \frac{-7+40 \Delta +100 \Delta ^2-280 \Delta ^3-210 \Delta ^4+840 
\Delta ^5-700 \Delta ^6+240 \Delta ^7-30 \Delta 
^8}{304819200}\lp
+\frac{(5-42 \Delta -63 \Delta ^2+420 \Delta ^3-525 
\Delta ^4+252 \Delta ^5-42 \Delta ^6) \kappa 
^2}{15240960}\lp
+\frac{(-1+20 \Delta -50 \Delta ^2+40 \Delta ^3-10 
\Delta ^4) \kappa ^4}{1451520}+\frac{(-5+12 \Delta -6 \Delta ^2) 
\kappa ^6}{2177280}-\frac{\kappa ^8}{10160640},
\ea
and so on. The costant $\mc C_{0}(\Delta, \kappa)$ is the usual half-sum of $\kappa$ shifted values of the constant in (\ref{D.5})
\ba
\mc C_{0}(\Delta,\kappa) &= \frac{1}{2} (-1+\Delta ) \log (2 \pi )+\frac{1}{2} (1-\Delta +\kappa 
) \log (\Gamma (\Delta -\kappa ))+\frac{1}{2} (1-\Delta -\kappa ) 
\log (\Gamma (\Delta +\kappa ))\lp
+\frac{1}{2} (\zeta ^{(1,0)}(-1,\Delta 
-\kappa )+\zeta ^{(1,0)}(-1,\Delta +\kappa )).
\ea
\fi

%A15
\iffa

\section{Fermion determinant in  $AdS_{3,\b}$}
\la{app:spinor}

The  determinant  of  the squared Dirac  operator in   $AdS_{3,\b }$
(with periodic  boundary  condition)  
can be computed starting from the spinor heat kernel and applying the method of images
as in \cite{David:2009xg}.
 Here, we review  a somewhat more direct derivation in \cite{Kakkar:2023gzu}. 
%  emphasizing the main steps and focusing on the 
%case of $AdS_{3}$ and a pure thermal quotient (no identification on the angular
%$S^{1}$).

We start from the Euclidean Dirac action in curved space
\ba
S &= -\int d^{3}\xi\, \sqrt{g}\ \overline\psi(\slashed{D}+M)\psi, \qquad
\slashed{D} = e\indices{_{a}^{m}}\gamma^{a}D_{m}, \qquad D_{m}=\partial_{m}+\frac{1}{4}\gamma^{ab}\omega_{ab,m}.
\ea
In  the case of  $\AdS_{3}$  with $\mathbb R^2$  boundary in Poincar\'e coordinates $\xi^{m}=(z, x, y)$ we have
\be
ds^{2} = \frac{1}{z^{2}}(dz^{2}+dx^{2}+dy^{2}), \qquad \qquad e^{a} = \frac{1}{z}(dz, dx, dy), \ \ \ \ \
\omega^{12}=\frac{1}{z}dx\,,  \ \  \omega^{13}=\frac{1}{z}dy, 
\ee
%From the Cartan equation $de^{a}+\omega^{ab}\wedge e^{b}=0$, we determine the spin connection 
%\ba & -\frac{1}{z^{2}}dz\wedge dx + \omega^{21}\wedge \frac{dz}{z}=0\quad\to\quad \omega^{12}=\frac{1}{z}dx,\\
%& -\frac{1}{z^{2}}dz\wedge dy + \omega^{31}\wedge \frac{dz}{z}=0\quad\to\quad \omega^{13}=\frac{1}{z}dy.
%\ea
%Choosing $(\gamma_{1},\gamma_{2},\gamma_{3}) = (\sigma_{3},\sigma_{1},\sigma_{2})$, the Dirac operator is then
so that  \ba
\slashed{D}  %&= z\gamma_{1}\partial_{1}+z\gamma_{2}(\partial_{2}+\frac{1}{2z}\gamma_{1}\gamma_{2})+z\gamma_{3}(\partial_{3}+\frac{1}{2z}\gamma_{1}\gamma_{3})
= \gamma_{1}(z\partial_{z}-1)+z\gamma_{2}\partial_{x}+z\gamma_{3}\partial_{y}.
\ea
The  spinor eigenvalue equations  are  
\ba
& \slashed{D}\psi = i\l\psi, \qquad \psi = z^{3/2}\begin{pmatrix} \psi_{+}(z) \\ \psi_{-}(z)\end{pmatrix}\, e^{ik_{x}x+ik_{y}y},
\\  
\psi_{+}'+(ik_{x}+k_{y})\psi_{-} &+\frac{1}{2z}(1-2i\l)\psi_{+}=0,   \qquad 
\psi_{-}'+(-ik_{x}+k_{y})\psi_{+}+\frac{1}{2z}(1+2i\l)\psi_{-}=0.
\ea
Separating them, we get 
\be
z^{2}\partial^{2}_{z}\psi_{\pm}+z\partial_{z}\psi_{\pm}-\Big[k^{2}z^{2}+(i\l\mp\frac{1}{2})^{2}\Big]\psi_{\pm}=0, \qquad k^{2}=k_{x}^{2}+k_{y}^{2}.
\ee
%Solving and imposing the Dirac equation, 
The unique solution bounded at infinity is (see appendix A.1 of  \cite{Kakkar:2023gzu})
\ba
&\Psi_{\l, \bm k}(\xi) = \langle\xi|\l,\bm k\rangle =  \psi_{\l,\bm{k}}(kz)
\,e^{ik_{x}x+k_{y}y}, \qquad\qquad  \psi_{\l, \bm{k}}(kz) = (kz)^{3/2}\begin{pmatrix}
K_{i\l-\frac{1}{2}}(kz) \\
\frac{k}{ik_{x}+k_{y}}K_{i\l+\frac{1}{2}}(kz)
\end{pmatrix} \ , \\
&
\langle \l', \bm{k}'|\l, \bm{k}\rangle = \int d^{3}\xi \sqrt{g}\Psi^{\dagger}_{\l',\bm k'}(\xi)\Psi_{\l,\bm k}(\xi) = (2\pi)^{2}\delta^{(2)}(\bm k-\bm k')\, k^{2}\,  \frac{\delta(\l-\l')}{   \mu(\l)   }, \\
&\mu(\l) = \frac{1}{\pi}\frac{1}{\Gamma(\frac{1}{2}+i\l)\Gamma(\frac{1}{2}-i\l)} = \frac{1}{\pi^{2}}\cosh(\pi \l),
\qquad \ \ \int_{-\infty}^{\infty} d\l\, \mu(\l) \int\frac{d^{2}\bm{k}}{k^{2}(2\pi)^{2}}|\l, \bm{k}\rangle\, \langle \l, \bm{k}|=1 \ .
\ea
%that implies the resolution of the identity
%\be 1 = \int_{-\infty}^{\infty} d\l\, \mu(\l) \int\frac{d^{2}\bm{k}}{k^{2}(2\pi)^{2}}|\l, \bm{k}\rangle\, \langle \l, \bm{k}|
%\ee
%where $\mc N$ is a normalization to be fixed later (and possibly divergent being morally the dimension of $\Z$).
%This is because 
The metric \rf{51} of Euclidean $\AdS_{3,\b}$ in global coordinates (here  we used 
 \be
ds^{2}_{\AdS_{3,\b} }=dx'^{2}+\sinh^{2}x'\,  d\vp^{2} +  \cosh^{2}x'\,  dy'^{2} 
\ee
transformed  into the  Poincar\'e coordinates is 
\be
ds^{2} = \frac{1}{z^{2}}(dz^{2}+dw d\bar w), \qquad  z = \frac{e^{y'}}{\cosh x'}, \qquad 
w=\tanh x'\, e^{y'+i\vp}, 
\ee 
 so that the  quotient $y'\to y'+\beta$  corresponds  to the  %becomes in Poincare' coordinates the 
 uniform rescaling
\be
(z,w,\bar w)\ \ \ \to \ \  q\,(z,w,\bar w),\qquad\qquad  q \equiv e^{-\beta}  \ .
\ee
As a result,  in  the case when $y$ is  compactified on $S^1_\b$ 
%For the thermal quotient (without rotation), we write 
the  spinor eigenfunctions  may be   written in the form  ($\mc N$ is a normalization   constant) 
% (notice that $\xi\to c\, \xi$ is an invariance of $\slashed{D}$) 
\be
\wh{\Psi}_{\l, \bm k}(\xi) = \frac{1}{\mc N}\sum_{n\in \Z}\Psi_{\l, \bm k}(e^{-n\beta}\xi) \ . 
\ee
Then we have 
\ba  &
\wh{\Psi}_{\l, \bm k}(\xi) = \langle \xi | \l, \bm{k}\rangle_{\beta}, \qquad \qquad |\l, \bm{k}\rangle_{\beta} = \frac{1}{\mc N}\sum_{n\in \Z} |\l, q^{n}\bm k\rangle.
\\   &
{}_{\beta}\langle \l', \bm{k}'|\l, \bm{k}\rangle_{\beta} = \frac{1}{\mc N^{2}}\sum_{n,n'\in \Z} (2\pi)^{2}\delta^{(2)}(q^{n}\bm k-q^{n'}\bm k')\, 
q^{2n}\,k^{2}\frac{\delta(\l-\l')}{\mu(\l)},
\\
%and one can easily check that the resolution of the identity takes the same form as before
%\be
&  \int_{-\infty}^{\infty} d\l\, \mu(\l) \int\frac{d^{2}\bm{k}}{k^{2}(2\pi)^{2}}|\l, \bm{k}\rangle_{\beta}\, _{\beta}\langle \l, \bm{k}|=1\ . 
\ea
%indeed
% \frac{1}{\mc N^{2}}\sum_{n,n'\in \Z}|\l, e^{-\beta(n'-n)}\bm{k}\rangle_{\beta}e^{2n\beta}\, 
%e^{-2n\beta}
%\ea
We can then compute  spectral traces in  coordinate space.\footnote{Since the spectrum is continuous, it is 
not convenient to  take traces in the $|\l,\bm{k}\rangle$ basis due to 
a $\delta^{(3)}(0)$ term  related to 
%to be identified with 
the regularized volume of $\AdS_{3}/\mathbb Z$.
%, but also with a $\l$ dependent factor due to the non-triviality of the wave functions.
} Using $\mu(\l)=\mu(-\l)$   we  get,  in particular, 
\ba
\la{F.22}
& \Tr\frac{1}{\slashed{D}+M} =\int d^{3}\xi\, \sqrt{g}\, \langle\xi|\frac{1}{\slashed{D}+M}|\xi\rangle 
= \int_{-\infty}^{\infty} d\l\, \mu(\l)\frac{M}{\l^{2}+M^{2}} \int\frac{d^{2}\bm{k}}{k^{2}(2\pi)^{2}} \int d^{3}\xi\, \sqrt{g}\, 
| \langle \xi | \l, \bm{k}\rangle_{\beta}|^{2}\ , 
\\   &
 \int d^{3}\xi\,  \sqrt{g}\, 
| \langle \xi | \l, \bm{k}\rangle_{\beta}|^{2} = \frac{1}{\mc N^{2}}\sum_{n,n'\in\Z}\int d^{3}\xi\sqrt{g}\langle \l,\bm{k}|q^{n}\xi\rangle\, 
\langle q^{n'}\xi|\l,\bm{k}\rangle \lp
= \frac{1}{\mc N^{2}}\sum_{n,n'\in\Z}\int_{0}^{\infty}\frac{dz}{z^{3}}\delta^{(2)}((q^{n}-q^{n'})\bm{k})\, \psi_{\l, \bm{k}}(q^{n}k z)\psi_{\l,\bm{k}}^{*}(q^{n'}kz) \lp
= \frac{1}{\mc N^{2}}\sum_{n,n'\in\Z}\int_{0}^{\infty}\frac{dz}{z^{3}}q^{2n'}\delta^{(2)}((q^{n}-q^{n'})\bm{k})\, \psi_{\l, \bm{k}}(q^{n-n'}k z)\psi_{\l,\bm{k}}^{*}(kz).
\ea
where in the last step we rescaled $z\to q^{-n'}z$. Using the symmetry between $n,n'$ and 
$
\frac{q^{2n'}}{|q^{n}-q^{n'}|^{2}} = \frac{1}{|1-q^{n-n'}|^{2}},
$
we obtain % (dropping the zero temperature term)  ???
\ba
 \int d^{3}\xi\, & \sqrt{g}\, 
| \langle \xi | \l, \bm{k}\rangle_{\beta}|^{2} = \frac{2}{\mc N}\sum_{n=1}^{\infty}\frac{1}{(1-q^{n})^{2}}\int_{0}^{\infty}\frac{dz}{z^{3}} \delta^{(2)}(\bm{k})
\psi_{\l,\bm{k}}(q^{n}kz)\psi_{\l,\bm{k}}^{*}(kz).
\ea
The small $k$ limit of the product of wave functions is\footnote{We use that $K_{\nu}(z)\sim\frac{1}{2}\Gamma(\nu)(z/2)^{-\nu}+\frac{1}{2}\Gamma(-\nu)(z/2)^{n}$ for small $z$.}
\be
\psi_{\l,\bm{k}}(kz)\psi_{\l,\bm{k}}^{*}(kz') \to  \frac{1}{2}\Gamma(\frac{1}{2}+i\l)\Gamma(\frac{1}{2}-i\l)(z^{i\l-\frac{1}{2}}z'^{-i\l-\frac{1}{2}}
+z^{-i\l-\frac{1}{2}}z'^{i\l-\frac{1}{2}})\,k^{2}(zz')^{3/2}.
\ee
Hence, 
\ba
 & \int d^{3}\xi\,  \sqrt{g}\, 
| \langle \xi | \l, \bm{k}\rangle_{\beta}|^{2} 
= \frac{2}{\mc N}\sum_{n=1}^{\infty}\frac{1}{(1-q^{n})^{2}}\int_{0}^{\infty}\frac{dz}{z^{3}}
\delta^{(2)}(\bm{k})\frac{1}{2}\Gamma(\frac{1}{2}+i\l)\Gamma(\frac{1}{2}-i\l)(kz)^{2}q^{3/2}(q^{n(i\l-\frac{1}{2})}+q^{n(-i\l-\frac{1}{2})})\lp
= k^{2}\delta^{(2)}(\bm{k})\, \int_{0}^{\infty}\frac{dz}{z}\  \frac{2}{\mc N}\sum_{n=1}^{\infty}\frac{q^{3/2}}{(1-q^{n})^{2}}
\frac{1}{2}\Gamma(\frac{1}{2}+i\l)\Gamma(\frac{1}{2}-i\l)q^{3/2}(q^{n(i\l-\frac{1}{2})}+q^{n(-i\l-\frac{1}{2})}).
\ea
Plugging this back into (\ref{F.22}) and integrating over  $\l$ gives
\ba\la{d22}
& \Tr\frac{1}{\slashed{D}+M} 
= \frac{2}{\mc N}\,\text{sign}(M) \sum_{n=1}^{\infty}\frac{e^{-n\beta(1+|M|)}}{(1-e^{-n\beta})^{2}}\int_{0}^{\infty}\frac{dz}{z}.
\ea
The divergent integral   may be  cancelled against  the $1/\mc N$ factor. 
This is seen  by splitting the integration infinite interval in copies of the fundamental region
% {\bf not sure I understand this ?????}
\be
\frac{1}{\mc N}\int_{0}^{\infty}\frac{dz}{z}= \frac{1}{\mc N}\sum_{n\in\Z}\int_{e^{-(n+1)\beta}}^{e^{-n\beta}}\frac{dz}{z} =\beta\ .
\ee
Finally, the logarithm of the determinant  may be 
 obtained  by integrating \rf{d22}   with respect to the mass parameter
\be
\log \det (\slashed{D}+M) = -\int_{M}^{\infty}dM' \, \Tr\frac{1}{\slashed{D}+M'} 
= -2\sum_{n=1}^{\infty}\frac{1}{n}\frac{q^{n\Delta}}{(1-q^{n})^{2}} \ , \ \ \ \ \qquad 
q=e^{-\beta}, \ \ \Delta = 1+|M| \  . 
\ee
 Apart from the factor 2 reflecting 2  spinor   components 
   and $-1$ from Fermi statistics, this has the same form  is in the 
 bosonic scalar case  assuming    proper   choice of  $\Delta(M)$.  % with the fermion mass.

\fi

\small 

\bibliography{BT-Biblio}

\providecommand{\href}[2]{#2}\begingroup\raggedright\begin{thebibliography}{10}

\bibitem{Maldacena:1997re}
J.~M. Maldacena, \emph{{The Large N limit of superconformal field theories and
  supergravity}},
  \href{https://doi.org/10.1023/A:1026654312961}{\emph{Int.J.Theor.Phys.}
  {\bfseries 38} (1999) 1113}
  [\href{https://arxiv.org/abs/hep-th/9711200}{{\ttfamily hep-th/9711200}}].

\bibitem{Aharony:1999ti}
O.~Aharony, S.~S. Gubser, J.~M. Maldacena, H.~Ooguri and Y.~Oz, \emph{{Large N
  field theories, string theory and gravity}},
  \href{https://doi.org/10.1016/S0370-1573(99)00083-6}{\emph{Phys.Rept.}
  {\bfseries 323} (2000) 183}
  [\href{https://arxiv.org/abs/hep-th/9905111}{{\ttfamily hep-th/9905111}}].

\bibitem{Gueven:1992hh}
R.~Gueven, \emph{{Black p-brane solutions of D = 11 supergravity theory}},
  \href{https://doi.org/10.1016/0370-2693(92)90540-K}{\emph{Phys. Lett.}
  {\bfseries B276} (1992) 49}.

\bibitem{Gibbons:1993sv}
G.~W. Gibbons and P.~K. Townsend, \emph{{Vacuum interpolation in supergravity
  via super p-branes}},
  \href{https://doi.org/10.1103/PhysRevLett.71.3754}{\emph{Phys. Rev. Lett.}
  {\bfseries 71} (1993) 3754}
  [\href{https://arxiv.org/abs/hep-th/9307049}{{\ttfamily hep-th/9307049}}].

\bibitem{Henningson:1998gx}
M.~Henningson and K.~Skenderis, \emph{{The Holographic Weyl anomaly}},
  \href{https://doi.org/10.1088/1126-6708/1998/07/023}{\emph{JHEP} {\bfseries
  9807} (1998) 023} [\href{https://arxiv.org/abs/hep-th/9806087}{{\ttfamily
  hep-th/9806087}}].

\bibitem{Tseytlin:2000sf}
A.~A. Tseytlin, \emph{{R4 terms in 11 dimensions and conformal anomaly of (2,0)
  theory}}, \href{https://doi.org/10.1016/S0550-3213(00)00380-1}{\emph{Nucl.
  Phys.} {\bfseries B584} (2000) 233}
  [\href{https://arxiv.org/abs/hep-th/0005072}{{\ttfamily hep-th/0005072}}].

\bibitem{Beccaria:2014qea}
M.~Beccaria, G.~Macorini and A.~A. Tseytlin, \emph{{Supergravity one-loop
  corrections on AdS$_7$ and AdS$_3$, higher spins and AdS/CFT}},
  \href{https://doi.org/10.1016/j.nuclphysb.2015.01.014}{\emph{Nucl.Phys.}
  {\bfseries B892} (2015) 211}
  [\href{https://arxiv.org/abs/1412.0489}{{\ttfamily 1412.0489}}].

\bibitem{Beem:2014kka}
C.~Beem, L.~Rastelli and B.~C. van Rees, \emph{{$ \mathcal{W} $ symmetry in six
  dimensions}}, \href{https://doi.org/10.1007/JHEP05(2015)017}{\emph{JHEP}
  {\bfseries 1505} (2015) 017}
  [\href{https://arxiv.org/abs/1404.1079}{{\ttfamily 1404.1079}}].

\bibitem{Ohmori:2014kda}
K.~Ohmori, H.~Shimizu, Y.~Tachikawa and K.~Yonekura, \emph{{Anomaly polynomial
  of general 6d SCFTs}}, \href{https://doi.org/10.1093/ptep/ptu140}{\emph{PTEP}
  {\bfseries 2014} (2014) 103B07}
  [\href{https://arxiv.org/abs/1408.5572}{{\ttfamily 1408.5572}}].

\bibitem{Drukker:2020dcz}
N.~Drukker, M.~Probst and M.~Tr\'epanier, \emph{{Surface operators in the 6d N
  = (2, 0) theory}}, \href{https://doi.org/10.1088/1751-8121/aba1b7}{\emph{J.
  Phys. A} {\bfseries 53} (2020) 365401}
  [\href{https://arxiv.org/abs/2003.12372}{{\ttfamily 2003.12372}}].

\bibitem{Drukker:2020swu}
N.~Drukker, S.~Giombi, A.~A. Tseytlin and X.~Zhou, \emph{{Defect CFT in the 6d
  (2,0) theory from M2 brane dynamics in AdS$_7 \times$S$^4$}},
  \href{https://doi.org/10.1007/JHEP07(2020)101}{\emph{JHEP} {\bfseries 07}
  (2020) 101} [\href{https://arxiv.org/abs/2004.04562}{{\ttfamily
  2004.04562}}].

\bibitem{Aharony:2008ug}
O.~Aharony, O.~Bergman, D.~L. Jafferis and J.~Maldacena,
  \emph{{${\mathcal{N}}\!=6$ Superconformal Chern-Simons-Matter Theories,
  M2-Branes and Their Gravity Duals}},
  \href{https://doi.org/10.1088/1126-6708/2008/10/091}{\emph{JHEP} {\bfseries
  10} (2008) 091} [\href{https://arxiv.org/abs/0806.1218}{{\ttfamily
  0806.1218}}].

\bibitem{Witten:1998zw}
E.~Witten, \emph{{Anti-de Sitter space, thermal phase transition, and
  confinement in gauge theories}}, {\emph{Adv. Theor. Math. Phys.} {\bfseries
  2} (1998) 505} [\href{https://arxiv.org/abs/hep-th/9803131}{{\ttfamily
  hep-th/9803131}}].

\bibitem{Seiberg:1997ax}
N.~Seiberg, \emph{{Notes on theories with 16 supercharges}},
  \href{https://doi.org/10.1016/S0920-5632(98)00128-5}{\emph{Nucl. Phys. Proc.
  Suppl.} {\bfseries 67} (1998) 158}
  [\href{https://arxiv.org/abs/hep-th/9705117}{{\ttfamily hep-th/9705117}}].

\bibitem{Douglas:2010iu}
M.~R. Douglas, \emph{{On D=5 Super Yang-Mills Theory and (2,0) Theory}},
  \href{https://doi.org/10.1007/JHEP02(2011)011}{\emph{JHEP} {\bfseries 02}
  (2011) 011} [\href{https://arxiv.org/abs/1012.2880}{{\ttfamily 1012.2880}}].

\bibitem{Lambert:2010iw}
N.~Lambert, C.~Papageorgakis and M.~Schmidt-Sommerfeld, \emph{{M5-Branes,
  D4-Branes and Quantum 5D Super-Yang-Mills}},
  \href{https://doi.org/10.1007/JHEP01(2011)083}{\emph{JHEP} {\bfseries 01}
  (2011) 083} [\href{https://arxiv.org/abs/1012.2882}{{\ttfamily 1012.2882}}].

\bibitem{Bobev:2018ugk}
N.~Bobev, P.~Bomans and F.~F. Gautason, \emph{{Spherical Branes}},
  \href{https://doi.org/10.1007/JHEP08(2018)029}{\emph{JHEP} {\bfseries 08}
  (2018) 029} [\href{https://arxiv.org/abs/1805.05338}{{\ttfamily
  1805.05338}}].

\bibitem{Mezei:2018url}
M.~Mezei, S.~S. Pufu and Y.~Wang, \emph{{Chern-Simons theory from M5-branes and
  calibrated M2-branes}},
  \href{https://doi.org/10.1007/JHEP08(2019)165}{\emph{JHEP} {\bfseries 08}
  (2019) 165} [\href{https://arxiv.org/abs/1812.07572}{{\ttfamily
  1812.07572}}].

\bibitem{Gautason:2021vfc}
F.~F. Gautason and V.~G.~M. Puletti, \emph{{Precision holography for 5D Super
  Yang-Mills}}, \href{https://doi.org/10.1007/JHEP03(2022)018}{\emph{JHEP}
  {\bfseries 03} (2022) 018}
  [\href{https://arxiv.org/abs/2111.15493}{{\ttfamily 2111.15493}}].

\bibitem{Gibbons:1986cq}
G.~W. Gibbons, \emph{{Quantized flux tubes in Einstein-Maxwell theory and
  noncompact internal spaces}},  in \emph{{22nd Winter School of Theoretical
  Physics: Fields and Geometry}}, pp.~597--615, 5, 1986.

\bibitem{Russo:1995ik}
J.~G. Russo and A.~A. Tseytlin, \emph{{Magnetic flux tube models in superstring
  theory}}, \href{https://doi.org/10.1016/0550-3213(95)00629-X}{\emph{Nucl.
  Phys. B} {\bfseries 461} (1996) 131}
  [\href{https://arxiv.org/abs/hep-th/9508068}{{\ttfamily hep-th/9508068}}].

\bibitem{Tseytlin:1995zv}
A.~A. Tseytlin, \emph{{Closed superstrings in magnetic field: Instabilities and
  supersymmetry breaking}},
  \href{https://doi.org/10.1016/0920-5632(96)00354-4}{\emph{Nucl. Phys. Proc.
  Suppl.} {\bfseries 49} (1996) 338}
  [\href{https://arxiv.org/abs/hep-th/9510041}{{\ttfamily hep-th/9510041}}].

\bibitem{Russo:1998xv}
J.~G. Russo and A.~A. Tseytlin, \emph{{Green-Schwarz superstring action in a
  curved magnetic Ramond-Ramond background}},
  \href{https://doi.org/10.1088/1126-6708/1998/04/014}{\emph{JHEP} {\bfseries
  04} (1998) 014} [\href{https://arxiv.org/abs/hep-th/9804076}{{\ttfamily
  hep-th/9804076}}].

\bibitem{Giombi:2023vzu}
S.~Giombi and A.~A. Tseytlin, \emph{{Wilson Loops at Large N and the Quantum
  M2-Brane}}, \href{https://doi.org/10.1103/PhysRevLett.130.201601}{\emph{Phys.
  Rev. Lett.} {\bfseries 130} (2023) 201601}
  [\href{https://arxiv.org/abs/2303.15207}{{\ttfamily 2303.15207}}].

\bibitem{Beccaria:2023ujc}
M.~Beccaria, S.~Giombi and A.~A. Tseytlin, \emph{{Instanton Contributions to
  the ABJM Free Energy from Quantum M2 Branes}},
  \href{https://arxiv.org/abs/2307.14112}{{\ttfamily 2307.14112}}.

\bibitem{Gautason:2023igo}
F.~F. Gautason, V.~G.~M. Puletti and J.~van Muiden, \emph{{Quantized strings
  and instantons in holography}},
  \href{https://doi.org/10.1007/JHEP08(2023)218}{\emph{JHEP} {\bfseries 08}
  (2023) 218} [\href{https://arxiv.org/abs/2304.12340}{{\ttfamily
  2304.12340}}].

\bibitem{Kim:2012ava}
H.-C. Kim and S.~Kim, \emph{{M5-Branes from Gauge Theories on the 5-Sphere}},
  \href{https://doi.org/10.1007/JHEP05(2013)144}{\emph{JHEP} {\bfseries 05}
  (2013) 144} [\href{https://arxiv.org/abs/1206.6339}{{\ttfamily 1206.6339}}].

\bibitem{Kim:2012qf}
H.-C. Kim, J.~Kim and S.~Kim, \emph{{Instantons on the 5-Sphere and
  M5-Branes}},  \href{https://arxiv.org/abs/1211.0144}{{\ttfamily 1211.0144}}.

\bibitem{Kim:2016usy}
S.~Kim and K.~Lee, \emph{{Indices for 6 Dimensional Superconformal Field
  Theories}}, \href{https://doi.org/10.1088/1751-8121/aa5cbf}{\emph{J. Phys. A}
  {\bfseries 50} (2017) 443017}
  [\href{https://arxiv.org/abs/1608.02969}{{\ttfamily 1608.02969}}].

\bibitem{Young:2011aa}
D.~Young, \emph{{Wilson Loops in Five-Dimensional Super-Yang-Mills}},
  \href{https://doi.org/10.1007/JHEP02(2012)052}{\emph{JHEP} {\bfseries 02}
  (2012) 052} [\href{https://arxiv.org/abs/1112.3309}{{\ttfamily 1112.3309}}].

\bibitem{Marino:2004eq}
M.~Mari\~no, \emph{{Les Houches Lectures on Matrix Models and Topological
  Strings}},  10, 2004, \href{https://arxiv.org/abs/hep-th/0410165}{{\ttfamily
  hep-th/0410165}}.

\bibitem{Kim:2013nva}
H.-C. Kim, S.~Kim, S.-S. Kim and K.~Lee, \emph{{The General M5-Brane
  Superconformal Index}},  \href{https://arxiv.org/abs/1307.7660}{{\ttfamily
  1307.7660}}.

\bibitem{Assel:2015nca}
B.~Assel, D.~Cassani, L.~Di~Pietro, Z.~Komargodski, J.~Lorenzen and
  D.~Martelli, \emph{{The Casimir Energy in Curved Space and its Supersymmetric
  Counterpart}}, \href{https://doi.org/10.1007/JHEP07(2015)043}{\emph{JHEP}
  {\bfseries 07} (2015) 043}
  [\href{https://arxiv.org/abs/1503.05537}{{\ttfamily 1503.05537}}].

\bibitem{Bobev:2015kza}
N.~Bobev, M.~Bullimore and H.-C. Kim, \emph{{Supersymmetric Casimir Energy and
  the Anomaly Polynomial}},
  \href{https://doi.org/10.1007/JHEP09(2015)142}{\emph{JHEP} {\bfseries 09}
  (2015) 142} [\href{https://arxiv.org/abs/1507.08553}{{\ttfamily
  1507.08553}}].

\bibitem{BenettiGenolini:2016qwm}
P.~Benetti~Genolini, D.~Cassani, D.~Martelli and J.~Sparks, \emph{{The
  Holographic Supersymmetric Casimir Energy}},
  \href{https://doi.org/10.1103/PhysRevD.95.021902}{\emph{Phys. Rev. D}
  {\bfseries 95} (2017) 021902}
  [\href{https://arxiv.org/abs/1606.02724}{{\ttfamily 1606.02724}}].

\bibitem{BenettiGenolini:2016tsn}
P.~Benetti~Genolini, D.~Cassani, D.~Martelli and J.~Sparks, \emph{{Holographic
  Renormalization and Supersymmetry}},
  \href{https://doi.org/10.1007/JHEP02(2017)132}{\emph{JHEP} {\bfseries 02}
  (2017) 132} [\href{https://arxiv.org/abs/1612.06761}{{\ttfamily
  1612.06761}}].

\bibitem{Bobev:2019bvq}
N.~Bobev, P.~Bomans, F.~F. Gautason, J.~A. Minahan and A.~Nedelin,
  \emph{{Supersymmetric Yang-Mills, Spherical Branes, and Precision
  Holography}}, \href{https://doi.org/10.1007/JHEP03(2020)047}{\emph{JHEP}
  {\bfseries 03} (2020) 047}
  [\href{https://arxiv.org/abs/1910.08555}{{\ttfamily 1910.08555}}].

\bibitem{Gunaydin:1984wc}
M.~Gunaydin, P.~van Nieuwenhuizen and N.~Warner, \emph{{General Construction of
  the Unitary Representations of Anti-de Sitter Superalgebras and the Spectrum
  of the $S^{4}$ Compactification of Eleven-dimensional Supergravity}},
  \href{https://doi.org/10.1016/0550-3213(85)90129-4}{\emph{Nucl.Phys.}
  {\bfseries B255} (1985) 63}.

\bibitem{Bhattacharya:2008zy}
J.~Bhattacharya, S.~Bhattacharyya, S.~Minwalla and S.~Raju, \emph{{Indices for
  Superconformal Field Theories in 3, 5 and 6 Dimensions}},
  \href{https://doi.org/10.1088/1126-6708/2008/02/064}{\emph{JHEP} {\bfseries
  02} (2008) 064} [\href{https://arxiv.org/abs/0801.1435}{{\ttfamily
  0801.1435}}].

\bibitem{Arai:2020uwd}
R.~Arai, S.~Fujiwara, Y.~Imamura, T.~Mori and D.~Yokoyama, \emph{{Finite-$N$
  corrections to the M-brane indices}},
  \href{https://doi.org/10.1007/JHEP11(2020)093}{\emph{JHEP} {\bfseries 11}
  (2020) 093} [\href{https://arxiv.org/abs/2007.05213}{{\ttfamily
  2007.05213}}].

\bibitem{Imamura:2022aua}
Y.~Imamura, \emph{{Analytic Continuation for Giant Gravitons}},
  \href{https://doi.org/10.1093/ptep/ptac127}{\emph{PTEP} {\bfseries 2022}
  (2022) 103B02} [\href{https://arxiv.org/abs/2205.14615}{{\ttfamily
  2205.14615}}].

\bibitem{Gaiotto:2021xce}
D.~Gaiotto and J.~H. Lee, \emph{{The Giant Graviton Expansion}},
  \href{https://arxiv.org/abs/2109.02545}{{\ttfamily 2109.02545}}.

\bibitem{Drukker:2000rr}
N.~Drukker and D.~J. Gross, \emph{{An Exact prediction of N=4 SUSYM theory for
  string theory}}, \href{https://doi.org/10.1063/1.1372177}{\emph{J. Math.
  Phys.} {\bfseries 42} (2001) 2896}
  [\href{https://arxiv.org/abs/hep-th/0010274}{{\ttfamily hep-th/0010274}}].

\bibitem{Witten:1988hf}
E.~Witten, \emph{{Quantum Field Theory and the Jones Polynomial}},
  \href{https://doi.org/10.1007/BF01217730}{\emph{Commun. Math. Phys.}
  {\bfseries 121} (1989) 351}.

\bibitem{Kapustin:2009kz}
A.~Kapustin, B.~Willett and I.~Yaakov, \emph{{Exact Results for Wilson Loops in
  Superconformal Chern-Simons Theories with Matter}},
  \href{https://doi.org/10.1007/JHEP03(2010)089}{\emph{JHEP} {\bfseries 03}
  (2010) 089} [\href{https://arxiv.org/abs/0909.4559}{{\ttfamily 0909.4559}}].

\bibitem{Bergshoeff:1987cm}
E.~Bergshoeff, E.~Sezgin and P.~K. Townsend, \emph{{Supermembranes and
  eleven-dimensional supergravity}},
  \href{https://doi.org/10.1016/0370-2693(87)91272-X}{\emph{Phys. Lett.}
  {\bfseries B189} (1987) 75}.

\bibitem{Duff:1987cs}
M.~J. Duff, T.~Inami, C.~N. Pope, E.~Sezgin and K.~S. Stelle,
  \emph{{Semiclassical Quantization of the Supermembrane}},
  \href{https://doi.org/10.1016/0550-3213(88)90316-1}{\emph{Nucl. Phys. B}
  {\bfseries 297} (1988) 515}.

\bibitem{Bergshoeff:1987qx}
E.~Bergshoeff, E.~Sezgin and P.~K. Townsend, \emph{{Properties of the
  Eleven-Dimensional Super Membrane Theory}},
  \href{https://doi.org/10.1016/0003-4916(88)90050-4}{\emph{Annals Phys.}
  {\bfseries 185} (1988) 330}.

\bibitem{Mezincescu:1987kj}
L.~Mezincescu, R.~I. Nepomechie and P.~van Nieuwenhuizen, \emph{{Do
  supermembranes contain massless particles?}},
  \href{https://doi.org/10.1016/0550-3213(88)90085-5}{\emph{Nucl. Phys. B}
  {\bfseries 309} (1988) 317}.

\bibitem{Forste:1999yj}
S.~Forste, \emph{{Membrany corrections to the string anti-string potential in
  M5-brane theory}},
  \href{https://doi.org/10.1088/1126-6708/1999/05/002}{\emph{JHEP} {\bfseries
  05} (1999) 002} [\href{https://arxiv.org/abs/hep-th/9902068}{{\ttfamily
  hep-th/9902068}}].

\bibitem{Bergshoeff:1987dh}
E.~Bergshoeff, M.~J. Duff, C.~N. Pope and E.~Sezgin, \emph{{Supersymmetric
  Supermembrane Vacua and Singletons}},
  \href{https://doi.org/10.1016/0370-2693(87)91465-1}{\emph{Phys. Lett. B}
  {\bfseries 199} (1987) 69}.

\bibitem{Bergshoeff:1988uc}
E.~Bergshoeff, M.~J. Duff, C.~N. Pope and E.~Sezgin, \emph{{Compactifications
  of the Eleven-Dimensional Supermembrane}},
  \href{https://doi.org/10.1016/0370-2693(89)91053-8}{\emph{Phys. Lett. B}
  {\bfseries 224} (1989) 71}.

\bibitem{Duff:1989ez}
M.~J. Duff, C.~N. Pope and E.~Sezgin, \emph{{A Stable Supermembrane Vacuum With
  a Discrete Spectrum}},
  \href{https://doi.org/10.1016/0370-2693(89)90575-3}{\emph{Phys. Lett. B}
  {\bfseries 225} (1989) 319}.

\bibitem{deWit:1998yu}
B.~de~Wit, K.~Peeters, J.~Plefka and A.~Sevrin, \emph{{The M theory two-brane
  in AdS$_4\times S^7$ and AdS$_7\times S^4$}},
  \href{https://doi.org/10.1016/S0370-2693(98)01340-9}{\emph{Phys. Lett.}
  {\bfseries B443} (1998) 153}
  [\href{https://arxiv.org/abs/hep-th/9808052}{{\ttfamily hep-th/9808052}}].

\bibitem{Claus:1998fh}
P.~Claus, \emph{{Super M-brane actions in AdS(4) $\times$ S7 and AdS(7)
  $\times$ S4}}, \href{https://doi.org/10.1103/PhysRevD.59.066003}{\emph{Phys.
  Rev. D} {\bfseries 59} (1999) 066003}
  [\href{https://arxiv.org/abs/hep-th/9809045}{{\ttfamily hep-th/9809045}}].

\bibitem{Pasti:1998tc}
P.~Pasti, D.~P. Sorokin and M.~Tonin, \emph{{On gauge fixed superbrane actions
  in AdS superbackgrounds}},
  \href{https://doi.org/10.1016/S0370-2693(98)01597-4}{\emph{Phys. Lett. B}
  {\bfseries 447} (1999) 251}
  [\href{https://arxiv.org/abs/hep-th/9809213}{{\ttfamily hep-th/9809213}}].

\bibitem{Tseytlin:1996hs}
A.~A. Tseytlin, \emph{{On dilaton dependence of type II superstring action}},
  \href{https://doi.org/10.1088/0264-9381/13/6/003}{\emph{Class. Quant. Grav.}
  {\bfseries 13} (1996) L81}
  [\href{https://arxiv.org/abs/hep-th/9601109}{{\ttfamily hep-th/9601109}}].

\bibitem{deWit:1998tk}
B.~de~Wit, K.~Peeters and J.~Plefka, \emph{{Superspace geometry for
  supermembrane backgrounds}},
  \href{https://doi.org/10.1016/S0550-3213(98)00445-3}{\emph{Nucl. Phys. B}
  {\bfseries 532} (1998) 99}
  [\href{https://arxiv.org/abs/hep-th/9803209}{{\ttfamily hep-th/9803209}}].

\bibitem{Harvey:1999as}
J.~A. Harvey and G.~W. Moore, \emph{{Superpotentials and membrane instantons}},
   \href{https://arxiv.org/abs/hep-th/9907026}{{\ttfamily hep-th/9907026}}.

\bibitem{Cremmer:1979up}
E.~Cremmer and B.~Julia, \emph{{The SO(8) Supergravity}},
  \href{https://doi.org/10.1016/0550-3213(79)90331-6}{\emph{Nucl. Phys. B}
  {\bfseries 159} (1979) 141}.

\bibitem{Drukker:2005kx}
N.~Drukker and B.~Fiol, \emph{{All-genus calculation of Wilson loops using
  D-branes}}, \href{https://doi.org/10.1088/1126-6708/2005/02/010}{\emph{JHEP}
  {\bfseries 02} (2005) 010}
  [\href{https://arxiv.org/abs/hep-th/0501109}{{\ttfamily hep-th/0501109}}].

\bibitem{Kalkkinen:2003gq}
J.~Kalkkinen and K.~S. Stelle, \emph{{Form field gauge symmetry in M theory}},
  \href{https://doi.org/10.1002/prop.200310108}{\emph{Fortsch. Phys.}
  {\bfseries 51} (2003) 856}
  [\href{https://arxiv.org/abs/hep-th/0302164}{{\ttfamily hep-th/0302164}}].

\bibitem{Duff:1987bx}
M.~J. Duff, P.~S. Howe, T.~Inami and K.~S. Stelle, \emph{{Superstrings in D=10
  from Supermembranes in D=11}},
  \href{https://doi.org/10.1016/0370-2693(87)91323-2}{\emph{Phys. Lett. B}
  {\bfseries 191} (1987) 70}.

\bibitem{Giombi:2020mhz}
S.~Giombi and A.~A. Tseytlin, \emph{{Strong coupling expansion of circular
  Wilson loops and string theories in AdS$_5 \times {\rm S}^5$ and AdS$_4
  \times {\rm CP}^3$}},
  \href{https://doi.org/10.1007/JHEP10(2020)130}{\emph{JHEP} {\bfseries 10}
  (2020) 130} [\href{https://arxiv.org/abs/2007.08512}{{\ttfamily
  2007.08512}}].

\bibitem{Minahan:2013jwa}
J.~A. Minahan, A.~Nedelin and M.~Zabzine, \emph{{5D super Yang-Mills theory and
  the correspondence to AdS$_7$/CFT$_6$}},
  \href{https://doi.org/10.1088/1751-8113/46/35/355401}{\emph{J. Phys. A}
  {\bfseries 46} (2013) 355401}
  [\href{https://arxiv.org/abs/1304.1016}{{\ttfamily 1304.1016}}].

\bibitem{Gibbons:2006ij}
G.~W. Gibbons, M.~J. Perry and C.~N. Pope, \emph{{Partition functions, the
  Bekenstein bound and temperature inversion in anti-de Sitter space and its
  conformal boundary}},
  \href{https://doi.org/10.1103/PhysRevD.74.084009}{\emph{Phys. Rev.}
  {\bfseries D74} (2006) 084009}
  [\href{https://arxiv.org/abs/hep-th/0606186}{{\ttfamily hep-th/0606186}}].

\bibitem{Giombi:2008vd}
S.~Giombi, A.~Maloney and X.~Yin, \emph{{One-loop Partition Functions of 3D
  Gravity}}, \href{https://doi.org/10.1088/1126-6708/2008/08/007}{\emph{JHEP}
  {\bfseries 0808} (2008) 007}
  [\href{https://arxiv.org/abs/0804.1773}{{\ttfamily 0804.1773}}].

\bibitem{Gopakumar:2011qs}
R.~Gopakumar, R.~K. Gupta and S.~Lal, \emph{{The Heat Kernel on $AdS$}},
  \href{https://doi.org/10.1007/JHEP11(2011)010}{\emph{JHEP} {\bfseries 1111}
  (2011) 010} [\href{https://arxiv.org/abs/1103.3627}{{\ttfamily 1103.3627}}].

\bibitem{Giombi:2014yra}
S.~Giombi, I.~R. Klebanov and A.~A. Tseytlin, \emph{{Partition Functions and
  Casimir Energies in Higher Spin $AdS_{d+1}/CFT_d$}},
  \href{https://doi.org/10.1103/PhysRevD.90.024048}{\emph{Phys. Rev.}
  {\bfseries D90} (2014) 024048}
  [\href{https://arxiv.org/abs/1402.5396}{{\ttfamily 1402.5396}}].

\bibitem{David:2009xg}
J.~R. David, M.~R. Gaberdiel and R.~Gopakumar, \emph{{The Heat Kernel on
  AdS$_{3}$ and its Applications}},
  \href{https://doi.org/10.1007/JHEP04(2010)125}{\emph{JHEP} {\bfseries 1004}
  (2010) 125} [\href{https://arxiv.org/abs/0911.5085}{{\ttfamily 0911.5085}}].

\bibitem{Datta:2011za}
S.~Datta and J.~R. David, \emph{{Higher Spin Quasinormal Modes and One-Loop
  Determinants in the BTZ Black Hole}},
  \href{https://doi.org/10.1007/JHEP03(2012)079}{\emph{JHEP} {\bfseries 03}
  (2012) 079} [\href{https://arxiv.org/abs/1112.4619}{{\ttfamily 1112.4619}}].

\bibitem{Datta:2012gc}
S.~Datta and J.~R. David, \emph{{Higher Spin Fermions in the BTZ Black Hole}},
  \href{https://doi.org/10.1007/JHEP07(2012)079}{\emph{JHEP} {\bfseries 07}
  (2012) 079} [\href{https://arxiv.org/abs/1202.5831}{{\ttfamily 1202.5831}}].

\bibitem{Kakkar:2023gzu}
A.~Kakkar and S.~Sarkar, \emph{{Phases of Theories with Fermions in AdS}},
  \href{https://doi.org/10.1007/JHEP06(2023)009}{\emph{JHEP} {\bfseries 06}
  (2023) 009} [\href{https://arxiv.org/abs/2303.02711}{{\ttfamily
  2303.02711}}].

\bibitem{Henningson:1998cd}
M.~Henningson and K.~Sfetsos, \emph{{Spinors and the AdS / CFT
  correspondence}},
  \href{https://doi.org/10.1016/S0370-2693(98)00559-0}{\emph{Phys. Lett.}
  {\bfseries B431} (1998) 63}
  [\href{https://arxiv.org/abs/hep-th/9803251}{{\ttfamily hep-th/9803251}}].

\bibitem{Gang:2011xp}
D.~Gang, E.~Koh, K.~Lee and J.~Park, \emph{{ABCD of 3d ${\cal N}=8$ and 4
  Superconformal Field Theories}},
  \href{https://arxiv.org/abs/1108.3647}{{\ttfamily 1108.3647}}.

\bibitem{Closset:2013vra}
C.~Closset, T.~T. Dumitrescu, G.~Festuccia and Z.~Komargodski, \emph{{The
  Geometry of Supersymmetric Partition Functions}},
  \href{https://doi.org/10.1007/JHEP01(2014)124}{\emph{JHEP} {\bfseries 01}
  (2014) 124} [\href{https://arxiv.org/abs/1309.5876}{{\ttfamily 1309.5876}}].

\bibitem{Cagnazzo:2009zh}
A.~Cagnazzo, D.~Sorokin and L.~Wulff, \emph{{String instanton in AdS(4) x
  CP3}}, \href{https://doi.org/10.1007/JHEP05(2010)009}{\emph{JHEP} {\bfseries
  05} (2010) 009} [\href{https://arxiv.org/abs/0911.5228}{{\ttfamily
  0911.5228}}].

\bibitem{Drukker:2011zy}
N.~Drukker, M.~Marino and P.~Putrov, \emph{{Nonperturbative aspects of ABJM
  theory}}, \href{https://doi.org/10.1007/JHEP11(2011)141}{\emph{JHEP}
  {\bfseries 11} (2011) 141} [\href{https://arxiv.org/abs/1103.4844}{{\ttfamily
  1103.4844}}].

\bibitem{Hatsuda:2013gj}
Y.~Hatsuda, S.~Moriyama and K.~Okuyama, \emph{{Instanton Bound States in ABJM
  Theory}}, \href{https://doi.org/10.1007/JHEP05(2013)054}{\emph{JHEP}
  {\bfseries 05} (2013) 054} [\href{https://arxiv.org/abs/1301.5184}{{\ttfamily
  1301.5184}}].

\bibitem{Sakaguchi:2010dg}
M.~Sakaguchi, H.~Shin and K.~Yoshida, \emph{{Semiclassical Analysis of M2-brane
  in $AdS_4 x S^7 / Z_k$}},
  \href{https://doi.org/10.1007/JHEP12(2010)012}{\emph{JHEP} {\bfseries 12}
  (2010) 012} [\href{https://arxiv.org/abs/1007.3354}{{\ttfamily 1007.3354}}].

\bibitem{Klemm:2012ii}
A.~Klemm, M.~Mari\~no, M.~Schiereck and M.~Soroush,
  \emph{{Aharony-Bergman-Jafferis--Maldacena Wilson Loops in the Fermi Gas
  Approach}}, \href{https://doi.org/10.5560/ZNA.2012-0118}{\emph{Z.
  Naturforsch. A} {\bfseries 68} (2013) 178}
  [\href{https://arxiv.org/abs/1207.0611}{{\ttfamily 1207.0611}}].

\bibitem{Nishioka:2008gz}
T.~Nishioka and T.~Takayanagi, \emph{{On Type IIA Penrose Limit and
  ${\mathcal{N}}\!=6$ Chern-Simons Theories}},
  \href{https://doi.org/10.1088/1126-6708/2008/08/001}{\emph{JHEP} {\bfseries
  08} (2008) 001} [\href{https://arxiv.org/abs/0806.3391}{{\ttfamily
  0806.3391}}].

\bibitem{Diaz:2007an}
D.~E. Diaz and H.~Dorn, \emph{{Partition functions and double-trace
  deformations in AdS/CFT}},
  \href{https://doi.org/10.1088/1126-6708/2007/05/046}{\emph{JHEP} {\bfseries
  0705} (2007) 046} [\href{https://arxiv.org/abs/hep-th/0702163}{{\ttfamily
  hep-th/0702163}}].

\bibitem{Kallen:2012zn}
J.~Kallen, J.~A. Minahan, A.~Nedelin and M.~Zabzine, \emph{{$N^3$-behavior from
  5D Yang-Mills theory}},
  \href{https://doi.org/10.1007/JHEP10(2012)184}{\emph{JHEP} {\bfseries 10}
  (2012) 184} [\href{https://arxiv.org/abs/1207.3763}{{\ttfamily 1207.3763}}].

\bibitem{Minahan:2016xwk}
J.~A. Minahan, \emph{{Matrix models for 5d super Yang\textendash{}Mills}},
  \href{https://doi.org/10.1088/1751-8121/aa5cbe}{\emph{J. Phys. A} {\bfseries
  50} (2017) 443015} [\href{https://arxiv.org/abs/1608.02967}{{\ttfamily
  1608.02967}}].

\bibitem{Cardy:1991kr}
J.~L. Cardy, \emph{{Operator content and modular properties of higher
  dimensional conformal field theories}},
  \href{https://doi.org/10.1016/0550-3213(91)90024-R}{\emph{Nucl. Phys.}
  {\bfseries B366} (1991) 403}.

\bibitem{Dunne:2007rt}
G.~V. Dunne, \emph{{Functional determinants in quantum field theory}},
  \href{https://doi.org/10.1088/1751-8113/41/30/304006}{\emph{J. Phys. A}
  {\bfseries 41} (2008) 304006}
  [\href{https://arxiv.org/abs/0711.1178}{{\ttfamily 0711.1178}}].

\bibitem{Kruczenski:2008zk}
M.~Kruczenski and A.~Tirziu, \emph{{Matching the circular Wilson loop with dual
  open string solution at 1-loop in strong coupling}},
  \href{https://doi.org/10.1088/1126-6708/2008/05/064}{\emph{JHEP} {\bfseries
  05} (2008) 064} [\href{https://arxiv.org/abs/0803.0315}{{\ttfamily
  0803.0315}}].

\bibitem{Aguilera-Damia:2018rjb}
J.~Aguilera-Damia, A.~Faraggi, L.~Pando~Zayas, V.~Rathee and G.~A. Silva,
  \emph{{Functional Determinants of Radial Operators in $AdS_2$}},
  \href{https://doi.org/10.1007/JHEP06(2018)007}{\emph{JHEP} {\bfseries 06}
  (2018) 007} [\href{https://arxiv.org/abs/1802.06789}{{\ttfamily
  1802.06789}}].

\bibitem{Denef:2009kn}
F.~Denef, S.~A. Hartnoll and S.~Sachdev, \emph{{Black Hole Determinants and
  Quasinormal Modes}},
  \href{https://doi.org/10.1088/0264-9381/27/12/125001}{\emph{Class. Quant.
  Grav.} {\bfseries 27} (2010) 125001}
  [\href{https://arxiv.org/abs/0908.2657}{{\ttfamily 0908.2657}}].

\end{thebibliography}\endgroup
\bibliographystyle{JHEP-v2.9}
\end{document}